\documentclass[]{emulateapj}

\usepackage{natbib}

\usepackage{amsmath}
\usepackage{epsfig}
\usepackage{graphicx}
\usepackage{amssymb}
\usepackage{color}
\usepackage{epstopdf}
\usepackage{rotating}
\newcommand{\eps}{e$^{-}/$s}
\newcommand{\nw}{nW/m$^2$/sr}

\shorttitle{CIBER: The Wide Field Imagers}
\shortauthors{BOCK ET AL. (CIBER COLLABORATION)}
\begin{document}

\slugcomment{Submitted to ApJS February 13 2012; accepted June 20 2012 as part of CIBER Instrument Special Issue.}

\title{The Cosmic Infrared Background Experiment (CIBER): The
Wide-Field Imagers}

\author{J.~Bock \altaffilmark{1,2}, I.~Sullivan \altaffilmark{3},
  T.~Arai\altaffilmark{4,5}, J. Battle\altaffilmark{1},
  A.~Cooray\altaffilmark{6}, V. Hristov\altaffilmark{2},
  B.~Keating\altaffilmark{7}, M.~G. Kim\altaffilmark{8},
  A.~C. Lam\altaffilmark{2},
  D.~H. Lee\altaffilmark{9}, L.~R. Levenson\altaffilmark{2},
  P.~Mason\altaffilmark{2}, T.~Matsumoto\altaffilmark{4,8,9},
  S.~Matsuura\altaffilmark{4} , K.~Mitchell-Wynne\altaffilmark{6},
  U.~W. Nam\altaffilmark{10} , T.~Renbarger\altaffilmark{7},
  J.~Smidt\altaffilmark{6}, K.~Suzuki\altaffilmark{11},
  K.~Tsumura\altaffilmark{4}, T.~Wada \altaffilmark{4},
  and M.~Zemcov\altaffilmark{2,1}}

\altaffiltext{1}{Jet Propulsion Laboratory (JPL), National Aeronautics and
Space Administration (NASA), Pasadena, CA 91109, USA}
\altaffiltext{2}{Department of Physics, Mathematics and Astronomy, California
Institute of Technology, Pasadena, CA 91125, USA}
\altaffiltext{3}{Department of Physics, The University of Washington, Seattle,
WA 98195, USA}
\altaffiltext{4}{Department of Space Astronomy and Astrophysics, Institute of Space
and Astronautical Science (ISAS), Japan Aerospace Exploration
Agency (JAXA), Sagamihara, Kanagawa 252-5210, Japan}
\altaffiltext{5}{Department of Physics, Graduate School of Science,
  The University of Tokyo, Tokyo 113-0033, Japan}
\altaffiltext{6}{Center for Cosmology, University of California, Irvine, Irvine,
CA 92697, USA}
\altaffiltext{7}{Department of Physics, University of California, San Diego,
San Diego, CA 92093, USA}
\altaffiltext{8}{Department of Physics and Astronomy, Seoul National University,
Seoul 151-742, Korea}
\altaffiltext{9}{Institute of Astronomy and Astrophysics, Academia
  Sinica, National Taiwan University, Taipei 10617, Taiwan R.~O.~C.}
\altaffiltext{10}{Korea Astronomy and Space Science Institute (KASI), Daejeon
305-348, Korea}
\altaffiltext{11}{Instrument Development Group of Technical Center, Nagoya
University, Nagoya, Aichi 464-8602, Japan}

\begin{abstract}
We have developed and characterized an imaging instrument to measure
the spatial properties of the diffuse near-infrared extragalactic
background light in a search for fluctuations from $z > 6$ galaxies during
the epoch of reionization.  The instrument is part of the Cosmic
Infrared Background Experiment (CIBER), designed to observe the
extragalactic background light above the Earth's atmosphere during a
suborbital sounding rocket flight.  The imaging instrument
incorporates a $2^{\circ} \times 2^{\circ}$ field of view, to measure
fluctuations over the predicted peak of the spatial power spectrum at
10 arcminutes, and $7'' \times 7''$ pixels, to remove lower redshift
galaxies to a depth sufficient to reduce the low-redshift galaxy
clustering foreground below instrumental sensitivity.  The imaging
instrument employs two cameras with $\Delta \lambda /
\lambda \sim 0.5$ bandpasses centered at $1.1 \, \mu$m and $1.6 \,
\mu$m to spectrally discriminate reionization extragalactic
background fluctuations from local foreground fluctuations.  CIBER
operates at wavelengths where the electromagnetic spectrum of the
reionization extragalactic background is thought to peak, and
complements fluctuations measurements by \textit{AKARI} and
\textit{Spitzer} at longer wavelengths.  We have characterized the
instrument in the laboratory, including measurements of the
sensitivity, flat-field response, stray light performance, and noise
properties.  Several modifications were made to the instrument
following a first flight in 2009 February.  The instrument performed
to specifications in subsequent flights in 2010 July and 2012 March, and
the scientific data are now being analyzed.
\vspace*{0.5cm}
\end{abstract}

\keywords{(Cosmology:) dark ages, reionization, first stars --
  (Cosmology:) diffuse radiation --
  Infrared: diffuse background --
  Instrumentation: photometers --
  space vehicles: instruments}

\section{Introduction}
\label{S:intro}

\setcounter{footnote}{0}

The extragalactic background light (EBL) is a measure of the
integrated radiation produced by stellar nucleosynthesis and
gravitational accretion over cosmic history.  The EBL must contain
the radiation produced during the epoch of reionization (the
reionization-EBL, or simply the REBL).  The REBL comes from
the UV and optical photons emitted by the first
ionizing stars and stellar remnants, radiation that is now redshifted
into the near-infrared (NIR).  The REBL is expected to peak at 1-2 um
due to the redshifted Lyman-$\alpha$ and Lyman-break features.  Furthermore
while the brightness of the REBL must be sufficient to initiate and sustain
ionization, the individual sources may be quite faint \citep{Salvaterra2011}.

We have developed a specialized imaging instrument to measure REBL
spatial fluctuations, consisting of two wide-field cameras that are part
of the Cosmic Infrared Background Experiment (CIBER; \citealt{Bock2006}),
developed to measure the absolute intensity, spectrum, and spatial
properties of the EBL.  CIBER's imaging cameras are combined with a
low-resolution spectrometer (LRS; \citealt{Tsumura2012}) designed to
measure the absolute sky brightness at wavelengths $0.75 < \lambda <
2.1 \, \mu$m, and a narrow-band spectrometer (NBS; \citealt{Korngut2012})
designed to measure the absolute ZL intensity using the $854.2 \,$nm Ca$\,$II
Fraunhofer line.  A full description of the CIBER payload, including the overall
mechanical and thermal design, and detailed descriptions of the focal plane housings,
calibration lamps, shutters, electronic systems, telemetry and data handling,
laboratory calibration equipment, flight events, and flight thermal performance,
is given in \citet{Zemcov2012}.  The observation sequence and science targets from
the first flight are available in \citet{Tsumura2010}.

In this paper, we describe the scientific background of EBL fluctuation measurements
in sections \ref{sS:science} and \ref{sS:drivers}, the instrument design in section
\ref{S:camera}, laboratory instrument characterization in section \ref{S:characterization},
modifications following the first flight in section \ref{S:mods}, and performance in
the second flight in section \ref{S:performance}.  Sensitivity calculations are given
in a short appendix.

\subsection{Science Background}
\label{sS:science}

Searching for the REBL appears to be more tractable in a multi-color
fluctuations measurement than by absolute photometry.  Absolute photometry,
measuring the sky brightness with a photometer and removing local foregrounds,
has proven to be problematic in the NIR, where the main difficulty is subtracting
the Zodiacal light (ZL) foreground, which is a combination of scattered sunlight and
thermal emission from interplanetary dust grains in our solar system.  However,
absolute photometry studies give consistent results in the far-infrared
(\citealt{Hauser1998}, \citealt{Fixsen1998}, \citealt{Juvela2009},
\citealt{Matsuura2011}, \citealt{Penin2011}).  These far-infrared
measurements are close to the EBL derived from galaxy counts though statistical
and lensing techniques that probe below the confusion limit (\citealt{Marsden2009},
\citealt{Zemcov2010}, \citealt{Bethermin2010}, \citealt{Berta2010}).
However in the NIR, at wavelengths appropriate for a REBL search, absolute EBL
measurements are not internally consistent (\citealt{Cambresy2001}, \citealt{Dwek1998},
\citealt{Matsumoto2005}, \citealt{Wright2001}, \citealt{Levenson2008}).  A
significant component of this disagreement is related to the choice of model used
to subtract ZL (\citealt{Kelsall1998}, \citealt{Wright2001}).  Furthermore, some
absolute EBL measurements (\citealt{Cambresy2001}, \citealt{Matsumoto2005}) are
significantly higher than the integrated galaxy light derived from source counts
(\citealt{Madau2000}, \citealt{Totani2001}, \citealt{Levenson2007}, \citealt{Keenan2010}).

The current disagreement between absolute measurements and galaxy counts are difficult
to reconcile with theoretical calculations \citep{Madau2005} or TeV absorption
measurements from blazars (\citealt{Gilmore2011}, \citealt{Aharonian2006},
\citealt{Schroedter2005}).  However TeV constraints on the NIR EBL require an
assumption about the intrinsic blazar spectrum \citep{Dwek2005}.  Furthermore cosmic
rays produced at the blazar are not attenuated by the EBL and can produce secondary
gamma rays that may explain the current TeV data without placing a serious constraint
on the NIR EBL \citep{Essey2010}.

\begin{figure*}[ht]
\epsfig{file=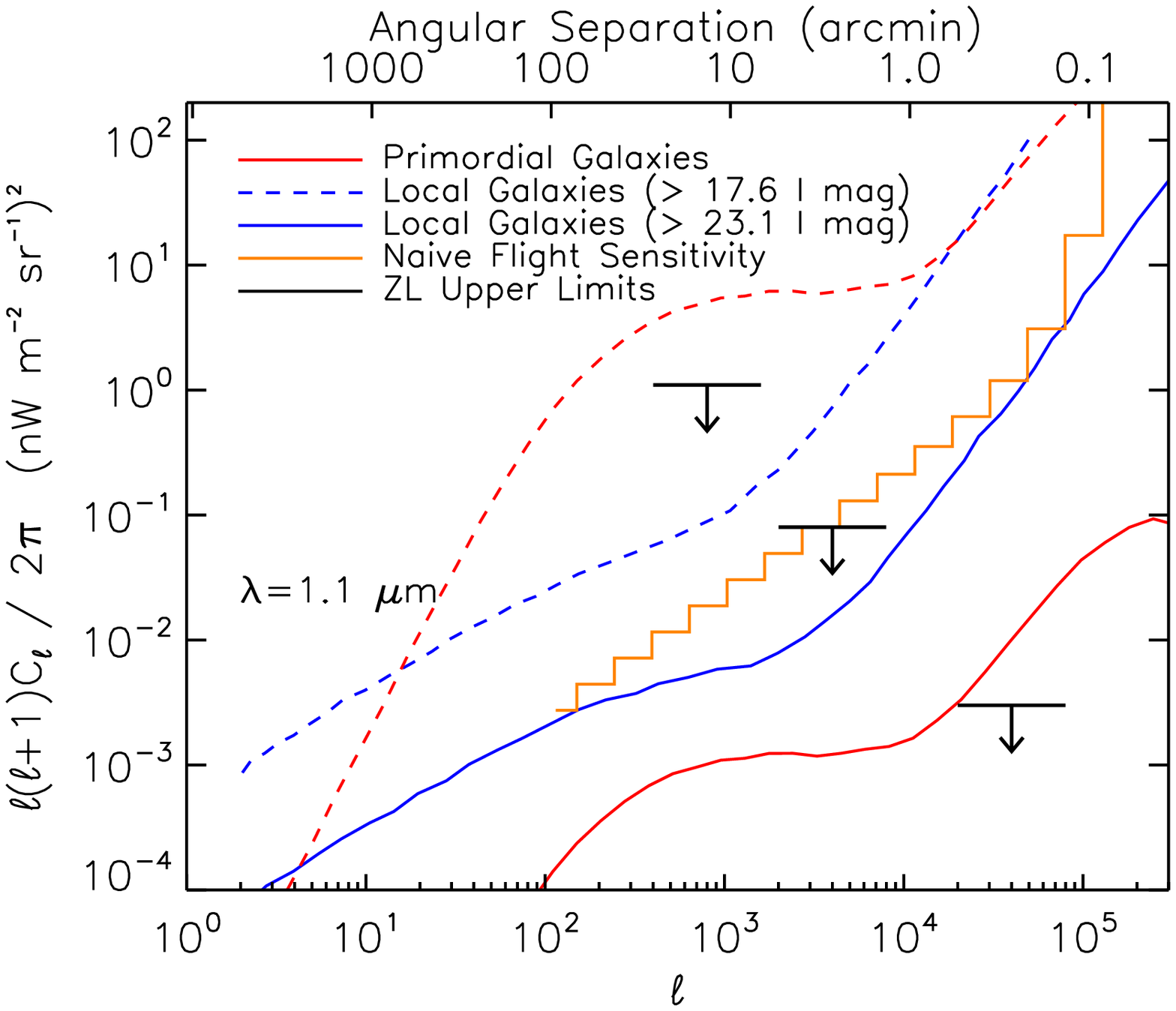,width=0.5\textwidth}
\epsfig{file=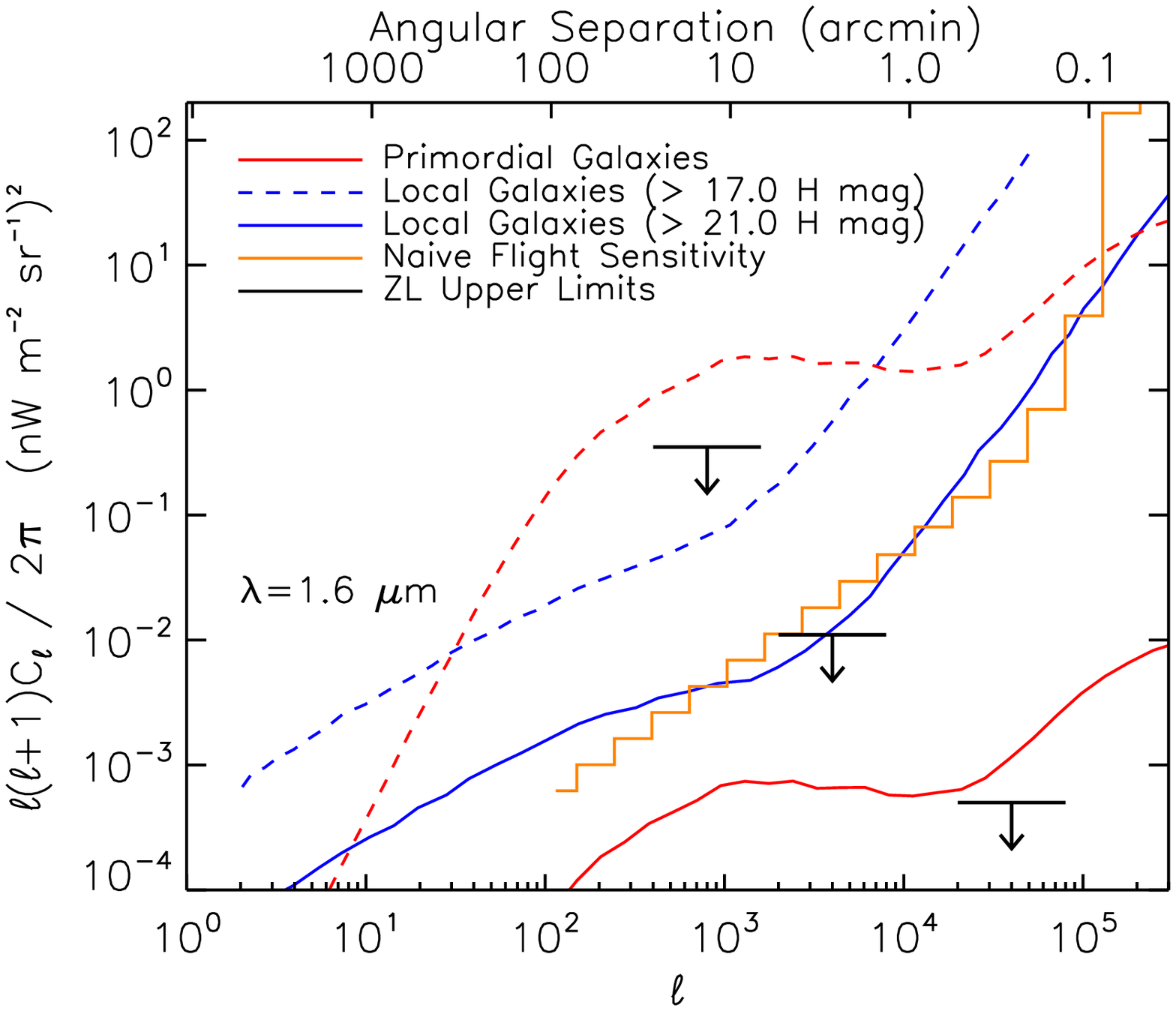,width=0.5\textwidth}
\caption{Power spectra of REBL and foreground fluctuations
  at $1.1 \, \mu$m (left) and $1.6 \, \mu$m (right).  In both cases
  the clustering power spectra of local ($z<3$) galaxies, for sources
  brighter than two different magnitude cutoffs, are shown as the blue
  solid and dashed lines.  These galaxy clustering power spectra are based
  on measured fluctuations as a function of cutoff magnitude from \citet{Sullivan2007}
  and are consistent with the predictions by \citet{Helgason2012} based on
  a large compilation of galaxy luminosity functions between $z=0$ and 4.
  The two red lines correspond to two expectations on the REBL
  anisotropy power spectrum as described in section \ref{sS:science}.  Upper
  limits to the ZL fluctuation power, shown in black, are scaled from experimental upper
  limits at longer wavelengths by the ZL spectrum.  The predicted CIBER sensitivities
  in both bands are shown in orange.  These are calculated with the instrument parameters
  listed in Table \ref{tab:imagerprops} assuming that the detector noise
  given in Table \ref{tab:imagersens} is uncorrelated and Gaussian over
  the array and using the $\Delta C_{\ell}$ formalism in \citet{Knox1995}.}
\label{fig:pwrspec}
\end{figure*}

Instead of measuring the absolute sky brightness, it is possible to
detect or constrain the REBL by studying the spatial properties of the
background (\citealt{Cooray2004}, \citealt{Kashlinsky2004}).  A spatial
power spectrum of the EBL contains a REBL clustering component, evident
at an angular scale of approximately 10 arcminutes as shown in Figure
\ref{fig:pwrspec}, that is related to the underlying power spectrum of
dark matter.  Numerical simulations of first galaxy formation indicate the
effects of non-linear clustering are significant \citep{Fernandez2010}.
There are also REBL fluctuations from the Poisson (unclustered shot noise)
component, but the amplitude of this term is more difficult to predict
as it is related to the number counts of the first galaxies, that is,
the brightness distribution and surface density of sources.  In
addition, REBL fluctuations are thought to have a characteristic
electromagnetic spectrum, peaking at the redshift-integrated
Lyman-$\alpha$ emission feature.  If reionization occurs at $z \sim
10$, this emission peak is redshifted into the NIR, with a spectral
shape that depends on the luminosity and duration of the epoch of
reionization.

Early measurements with the Diffuse Infrared Background Experiment
(DIRBE; \citealt{Kashlinsky2000}) and the Infrared Telescope in Space
(IRTS; \citealt{Matsumoto2005}) used fluctuations as a tracer of the
total EBL.  A first detection of REBL fluctuations was reported by
\citet{Kashlinsky2005} using the \textit{Spitzer} Infrared Array
Camera (IRAC; \citealt{Fazio2004}) in the 3.6 and $4.5 \, \mu$m bands
in $5 \times 5$ arcminute regions, corresponding to the IRAC field of
view.  The authors observe a departure from Poisson noise on $1{-}5$
arcminute scales which they attribute to first-light galaxies, after
ruling out Zodiacal, Galactic, and galaxy clustering foregrounds.
The observed brightness of the fluctuations is approximately constant
at 3.6 and $4.5 \, \mu$m.  This analysis was later extended to $10 \times 10
\,$arcmin fields, giving similar results \citep{Kashlinsky2007}.
\citet{Thompson2007a} studied a $144 \times 144 \,$arcsec field with
the Hubble Space Telescope (\textit{HST}) at 1.1 and $1.6 \, \mu$m,
finding no evidence for $z > 8$ galaxies contributing to the
\textit{HST} or the \textit{Spitzer} fluctuations
\citep{Thompson2007b}.  Finally, \citet{Matsumoto2011} report
first-light galaxy fluctuations with \textit{AKARI} at 2.4, 3.2 and
$4.1 \, \mu$m in a 10 arcminute field.  Their reported spectrum shows
a strong increase from 4.1 to $2.4 \, \mu$m, consistent with a
Rayleigh-Jeans spectrum.

In Figure \ref{fig:pwrspec} we show two predictions related to the angular power spectrum
of REBL anisotropies.  The lower prediction (solid red line) is from \citet{Cooray2012},
derived from the observed luminosity functions of Lyman dropout galaxies at redshifts
of 6, 7 and 8 \citep{Bouwens2008} at the bright end.  The reionization history involves
an optical depth to electron scattering of 0.09, consistent with
the WMAP 7-year measurement of $\tau=0.088 \pm 0.014$ \citep{Komatsu2011}. The absolute REBL
background is $0.3 \, nW m^{-2} sr^{-1}$ at 3.6 $\mu$m for this model.  \citet{Cooray2012}
improved on previous predictions \citep{Cooray2004} by accounting for non-linear clustering at
small angular scales with a halo model for reionization galaxies at $z > 6$.  Note that the
REBL fluctuation power is similar at 1.6 and 1.1 $\mu$m given the redshift
of reionization is around at $z \sim 10$.

The upper prediction (dashed red line) is normalized to the anisotropy amplitude level
reported by {\it Spitzer}-IRAC at 3.6 $\mu$m \citep{Kashlinsky2005}.  This power spectrum
requires an absolute REBL background between 2 to 3 $nW m^{-2} sr^{-1}$ at 3.6 $\mu$m.  We
scale the power spectra to shorter wavelengths based on a Rayleigh-Jeans spectrum, consistent
with the combined measurements of {\it Spitzer} and {\it AKARI} \citep{Matsumoto2011}.

Fluctuation measurements are only feasible if the contributions from
foregrounds can be removed.  Fortunately, it appears easier to remove
foregrounds in fluctuation measurements than in absolute
photometry measurements.  The largest foreground, ZL, is known to be
spatially uniform on spatial scales smaller than a degree
(\citealt{Abraham1997}, \citealt{Kashlinsky2005}, \citealt{Pyo2011}).
Furthermore, any spatial variations in ZL can be monitored and removed
by observing a field over a period of time, as the view through the
interplanetary dust cloud changes annually.  Galaxies and stars give
spatial fluctuations from Poisson variations and clustering.  These
can be eliminated by masking sources from the image, either through
detection or by using an external catalog of known sources.  Galaxy
clustering, arguably the most serious of these potential contaminants,
requires a sufficiently deep source cutoff to reduce the clustering
spectrum below the level of REBL fluctuations by masking sources.

\subsection{Theoretical Design Drivers}
\label{sS:drivers}

These early fluctuation results call for a next generation of improved
measurements at shorter wavelengths, spanning the expected peak of the
REBL electromagnetic spectrum, with wide angular coverage, to definitively
measure the expected peak in the REBL spatial power spectrum.  In order to
make a definitive REBL fluctuations measurement, we require: (1) a wide field
of view to allow measurements of the characteristic REBL spatial power spectrum,
(2) observations in multiple NIR bands in order to characterize the REBL
electromagnetic spectrum and distinguish it from potential foregrounds,
and (3) arcsecond angular resolution to remove galaxies to a sufficient depth
to minimize the galaxy clustering foreground signal.

High-fidelity spatial imaging on degree scales is problematic in the
NIR due to airglow emission from the Earth's atmosphere, which is some
$200 {-} 1500$ times brighter than the astrophysical sky in the NIR J,
H and K bands \citep{Allen1976}.  Airglow emission has time-variable
structure \citep{Ramsay1992} with spatial variations that increase on
larger angular scales, especially from $1^{\circ}$ to $10^{\circ}$
\citep{Adams1996}.  We therefore conduct observations on a sounding
rocket flight, at altitudes above the layers in the atmosphere
responsible for airglow emission at characteristic altitudes of $\sim
100 \,$km.  To measure the $\sim 10'$ peak in the REBL spatial power
spectrum, it is necessary to image an area of sky on the order of a
square degree.  While one can image a large field with a mosaic
using a small field of view, this requires a highly stable instrument.
A wide field of view allows a measurement using single exposures in
the short time available on a sounding rocket flight.

The REBL electromagnetic spectrum is predicted to peak at $1{-}2 \,
\mu$m (\citealt{Cooray2004}, \citealt{Kashlinsky2004}) due to the
redshift-integrated Lyman-$\alpha$ emission feature, with a decreasing
spectrum at longer wavelengths that depends on the history of
reionization and the presence of free-free emission from ionized gas
surrounding the first galaxies. Observations in the optical and
near-IR should detect this spectrum, which is distinct from that of
local foregrounds, namely ZL, stars, galaxies, scattered starlight
(i.e.~diffuse galactic light), and other Galactic emission.  Though
ideally the wavelength coverage would extend out to $\sim 5 \, \mu$m,
the key wavelengths for REBL science bracket the $1{-}2 \, \mu$m peak.
Longer wavelength information can be obtained by cross-correlating CIBER
data with overlapping wide-field \textit{Spitzer} and \textit{AKARI} maps.

The local-galaxy fluctuations foreground is mitigated by masking galaxies
down to a given flux threshold.  The masking depth needed depends on the
residual clustering and Poisson fluctuations of galaxies below the
cutoff flux.  \citet{Sullivan2007} measured galaxy clustering as a
function of cutoff from a wide-field ground-based NIR survey catalog.
We note that the REBL is best discriminated from low-redshift galaxy
clustering and Poisson fluctuations at ~10 arcminutes, as is evident
in Figure \ref{fig:pwrspec} by comparing the REBL and galaxy
clustering power spectra.  Thus wide-field observations are also helpful
for discriminating REBL from local galaxy fluctuations.

The flux cutoff needed to separate the optimistic REBL model from local galaxy
fluctuations is $\sim 17^{th}$ Vega magnitude at $1.6 \, \mu$m, as is evident from
the curves in Figure \ref{fig:pwrspec}.  The spatial density of galaxies brighter
than $17^{th}$ Vega magnitude is $N(>S) = 500$ galaxies per square
degree.  The cutoff required to remove galaxies well below the expected CIBER instrument
sensitivity is $\sim 23^{rd}$ Vega magnitude at $1.6 \, \mu$m, corresponding
to $N(>S) = 1.5 \times 10^{5}$ galaxies per square degree.  Thus we find an angular
resolution of $4 {-} 80$ arcseconds is needed to remove galaxies in order to lose
less than 25 \% of the pixels from masking.

Galaxy masking can be accomplished using ancillary observations with greater point
source depth, masking pixels in the CIBER images below the CIBER point source sensitivity.
The fields observed in the first two flights of CIBER, listed in Table \ref{tab:ancfields},
allows source masking using deep companion catalogs obtained in ground based NIR observations.
Details on first flight observations of these fields is available in \citet{Tsumura2010}.
These fields have also been observed in a search for REBL fluctuations by \textit{AKARI} and
\textit{Spitzer} at longer wavelengths, allowing for a cross-correlation analysis with CIBER.

\begin{table*}[htb]
\centering
\caption{CIBER Survey Fields and Ancillary Data Depths.}
\begin{tabular}{llccccl}
\hline
CIBER Field & Ancillary & $\lambda$ & Field Coverage &
\multicolumn{2}{c}{Ancillary Depth} & Reference \\
 & Coverage &  ($\mu$m) & (\%) & (Vega mag) & ($\sigma$) & \\ \hline
Bo\"{o}tes & NDWFS & 0.83 & 100 & 25.5 & 5 & \citet{Jannuzi1999} \\
 & NEWFIRM & 1.0 & 100 & 22.0 & 5 & \citet{Gonzalez2011} \\
 & NEWFIRM & 1.6 & 100 & 20.8 & 5 & \citet{Gonzalez2011} \\
 & NEWFIRM & 2.4 & 100 & 19.5 & 5 & \citet{Gonzalez2011} \\
 & \textit{Spitzer}-SDWFS & 3.6 & 100 & 19.7 & 5 & \citet{Ashby2009} \\
North Ecliptic Pole & Maidanak & 0.9 & 60 & 21.9 & 5 & \citet{Jeon2010} \\
 & CFHT & 1.2 & 50 & 24 & 4 & \citet{Hwang2007} \\
 & 2MASS & 1.6 & 100 & 17.9 & 10 & \citet{Cutri2003} \\
 & \textit{AKARI} & 2.4 & 98 & 19.7 & 5 & \citet{Lee2009} \\
ELIAS-N1 & UKIDSS-DR6 & 0.9 & 75 & 22.3 & 5 & \citet{Lawrence2007} \\
 & INT & 0.9 & 100 & 21.9 & 5 & \citet{GS2011} \\
 & 2MASS & 1.6 & 100 & 17.8 & 10 & \citet{Cutri2003} \\
 & \textit{Spitzer}-SWIRE & 3.6 & 100 & 18.6 & 10 & \citet{Lonsdale2003} \\
\hline
\end{tabular}
\label{tab:ancfields}
\end{table*}

\section{Instrument Design}
\label{S:camera}

\begin{figure*}[htb]
\centering
\epsfig{file=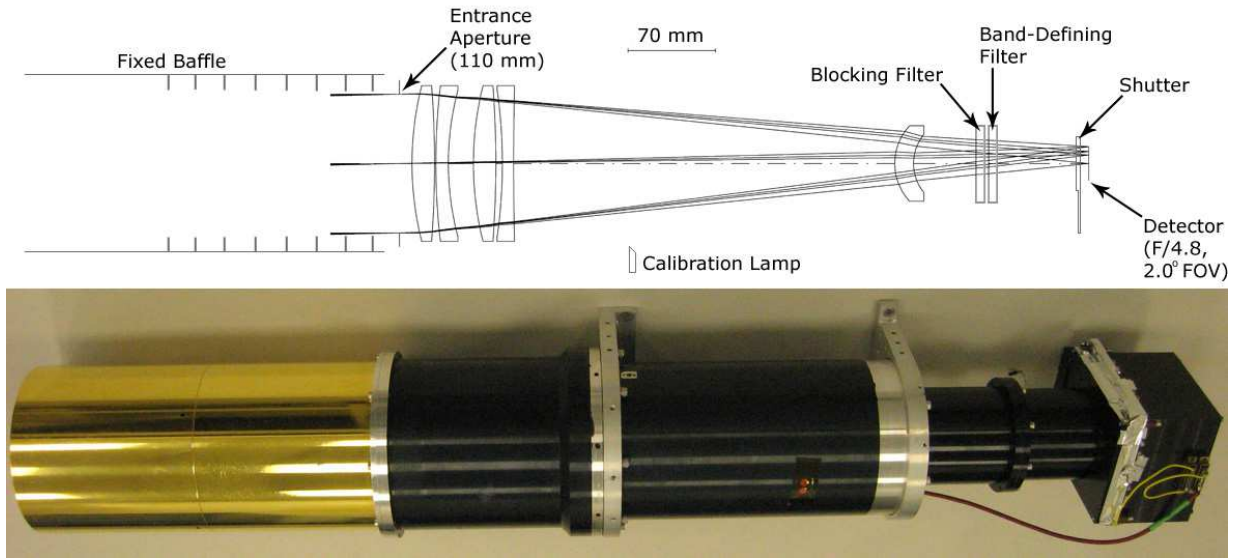,width=0.9\textwidth}
\caption{Schematic and photograph of the CIBER imaging camera.  Light
  enters the optical system at left and is imaged to the focal plane
  at right.  A fixed baffle is used to reduce scattering on the first
  optic.  The Imager assembly employs a fiber-fed calibration lamp
  system, band-defining and blocking filters, and a focal plane
  assembly as described in \citet{Zemcov2012}.  Both Imager assemblies
  used in CIBER are identical except for their band defining filters,
  set with $\Delta \lambda / \lambda \sim 0.5$ bandpasses centered at
  $1.1 \, \mu$m and $1.6 \, \mu$m, roughly corresponding to
  astronomical I and H band.  The photograph shows a fully assembled
  Imager in the lab.  The entire assembly mounts to the CIBER optical
  bench when installed in the payload and operates at $\sim 80 \,$K.}
\label{fig:imageroptics}
\end{figure*}

The Imager instrument consists of two wide-field refracting NIR
telescopes each with an $11 \,$cm aperture, combined with
band-defining filters, a cold shutter, and a $1024 \times 1024$ HgCdTe
$2.5 \, \mu$m Hawaii-1\footnote{Manufactured by Teledyne Scientific \&
  Imaging, LLC.} focal plane array.  The Imager optics were designed
and built by Genesia Corporation using the cryogenic index of
refraction measurements of \citet{Yamamuro2006}.  A schematic of the
assembly is shown in Figure \ref{fig:imageroptics}.  The assembly
housing the Imager optics are constructed from aluminum alloy 6061,
and the lenses are made from anti-reflection coated Silica,
S-FPL53 and S-TIL25 glass.  The assembly is carefully designed to maintain
optical alignment and focus through launch acceleration and vibration.
The aluminum housing is hard black anodized to reduce reflections
inside the cryogenic insert and telescope assembly, with the exception
of the static baffle at the front of the assembly which is gold plated
on its external surface and Epner laser black coated\footnote{This is
  a proprietary process of Epner Technology, Inc.} on its inner
surface.  This scheme serves to reduce the absorptivity of the baffle
on the side facing warm components at the front of the payload
section, and increase the absorptivity to NIR light on the inside.
At the other end of the camera, a focal plane assembly is mounted to
the back of the optical assembly and thermally isolated using Vespel
SP-1 standoffs.  The assembly includes a cold shutter and active
thermal control for each detector.  In addition, a calibration lamp
system illuminates the focal plane in a repeatable way to provide a
transfer standard during flight.  The design of the calibration lamp
system is common to all of the CIBER instruments and is presented in
\citet{Zemcov2012}.

\begin{figure}[htb]
\epsfig{file=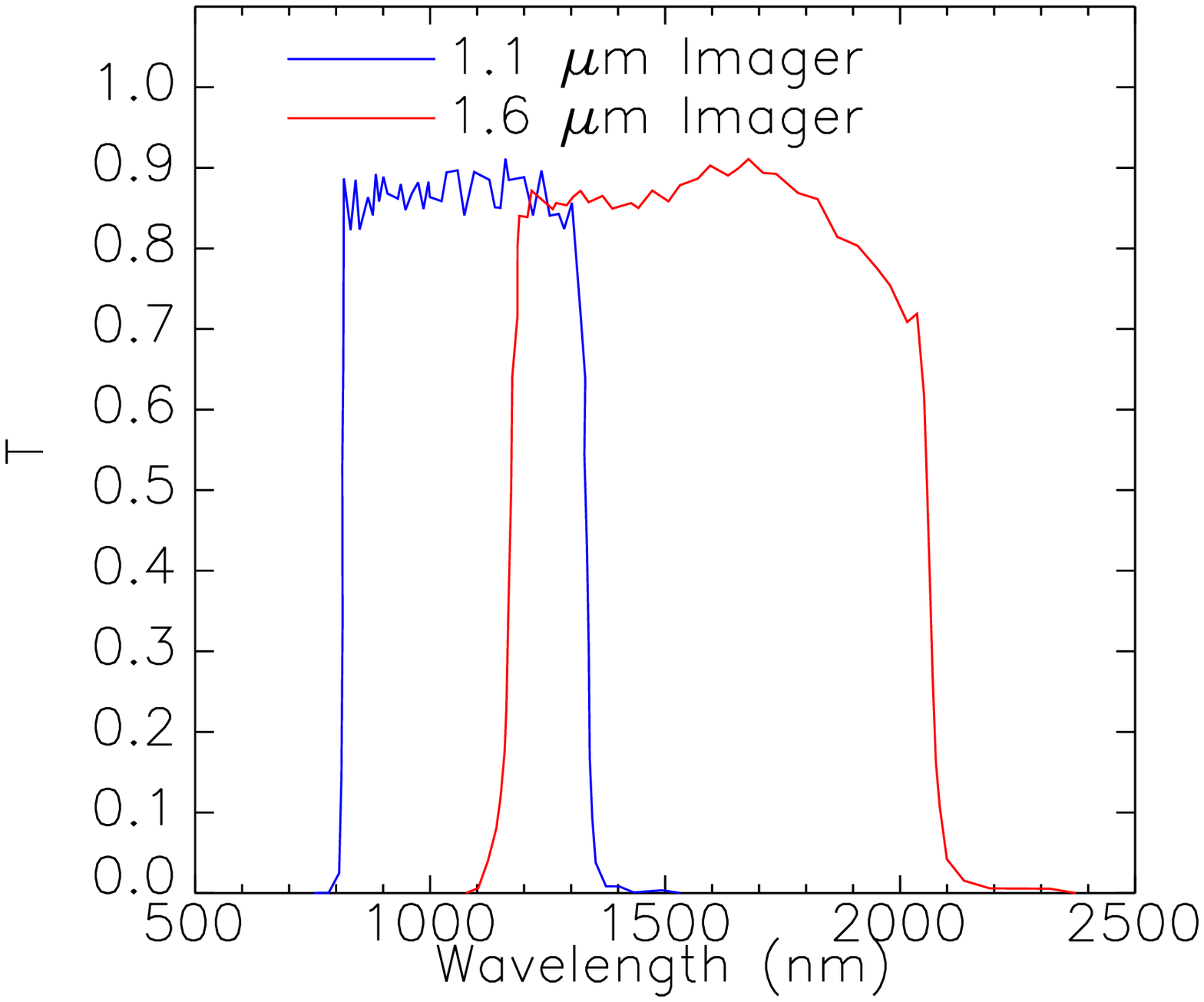,width=0.48\textwidth}
\caption{$1.1 \, \mu$m and $1.6 \, \mu$m Imager filter responses.
  These curves represent the transmission of the optical stack which
  includes band defining and blocking filters as well as 5 anti-reflection
  coated lenses.  This response does not include the response of the detector
  array, which typically cuts off at $\sim 900 \,$nm for a Hawaii-1
  array with a sapphire substrate (Mark Farris, private
  communication).}
\label{fig:filters}
\end{figure}

The optical transmittance of the two Imager filters are shown in
Figure \ref{fig:filters}.  The filter stack is located behind the
optical elements and in front of the focal plane assembly and cold
shutter as shown in Figure \ref{fig:imageroptics}.  Each lens provides
additional filtering for wavelengths that are out of band for both
instruments, as their anti-reflection coatings transmit less than
1.5\% of light with wavelengths shorter than $0.75 \, \mu$m or longer
than $2.0 \, \mu$m.

Table \ref{tab:imagerprops} summarizes the design properties of the
optics and detector system, and the measured efficiencies, bands, and
read noise for the two cameras.  The optical efficiency is the product
of the reflectance and absorption of the anti-reflection coated lenses
taken from witness samples.  The instrument performance is calculated
in the appendix based on data from Table \ref{tab:imagerprops} and
presented in Table \ref{tab:imagersens}.

\begin{table}[ht]
\centering
\caption{Imager Instrument Properties.}
\begin{tabular}{lccc}
\hline
 & 1.1 $\mu$m Band & 1.6 $\mu$m Band & Units \\
\hline
Wavelength Range & $900{-}1320 \ast$ & $1150{-}2040$ & nm\\
Pupil Diameter & 110 & 110 & mm \\
F\# & 4.95 & 4.95 \\
Focal Length & 545 & 545 & mm \\
Pixel Size & $7 \times 7$ & $7 \times 7$ & arcsec\\
Field of View & $2.0 \times 2.0$ & $2.0 \times 2.0$ & deg\\
Optics Efficiency & 0.90 & 0.90 & \\
Filter Efficiency & 0.92 & 0.89 & \\
Array QE & 0.51 & 0.70 & $\ast \ast$ \\
Total Efficiency & 0.42 & 0.56 & \\
Array Format & $1024^{2}$ & $1024^{2}$ \\
Pixel Pitch & 18 & 18 & $\mu$m\\
Read Noise (CDS) & 10 & 9 & e$^{-}$\\
Frame Interval & 1.78 & 1.78 & s\\
\hline
\multicolumn{4}{l}{$\ast$ We assume a $900 \,$nm cut-on wavelength
  from the} \\
\multicolumn{4}{l}{Hawaii-1 substrate.} \\
\multicolumn{4}{l}{$\ast \ast$ Array QE is estimated from QE measured at
  $2.2 \, \mu$m} \\
\multicolumn{4}{l}{for each array and scaled based on the response of}
\\
\multicolumn{4}{l}{a typical Hawaii-1.}  \\
\end{tabular}
\label{tab:imagerprops}
\vspace{5pt}
\end{table}

Once assembled, the cameras mount to an optical bench shared with the
LRS and NBS.  The completed instrument section is then inserted into
the experiment vacuum skin.  Like the other CIBER instruments, the
Imager optics are cooled to $\sim 80 \,$K to reduce their in-band
emission using a liquid nitrogen cryostat system.
\citet{Zemcov2012} describes the various payload configurations used
in calibration and in flight which allow both dark and optical testing in
the laboratory.

\section{Instrument Characterization}
\label{S:characterization}

REBL fluctuation measurements place demanding requirements on the
instrument, including the detector noise properties, linearity and
transient response, optical focus, control of stray radiation, and
knowledge of the flat field response.  We have carried out a series of
laboratory measurements to characterize these properties.

\subsection{Dark Current}
\label{sS:darkcurrent}

The detector dark current is measured in both flight and laboratory
configurations by closing the cold shutters, which attenuate the optical
signal by a measured factor of $\sim 10^{3}$.  Array data are acquired
at $6.8 \, \mu$s per pixel sample, so that the full array is read in
$1.78 \,$s.  The pixels are read non-destructively, and integrate
charge until reset.  The integration time may be selected, but the
flight integrations are typically $\sim 50\,$s.  To maximize the
signal-to-noise ratio, for each pixel we fit the measured output voltage
to a slope and an offset as described in \citet{Garnett1993}.  All CIBER
Imager data are analyzed using this method, except where noted.

\begin{figure}[htb]
\epsfig{file=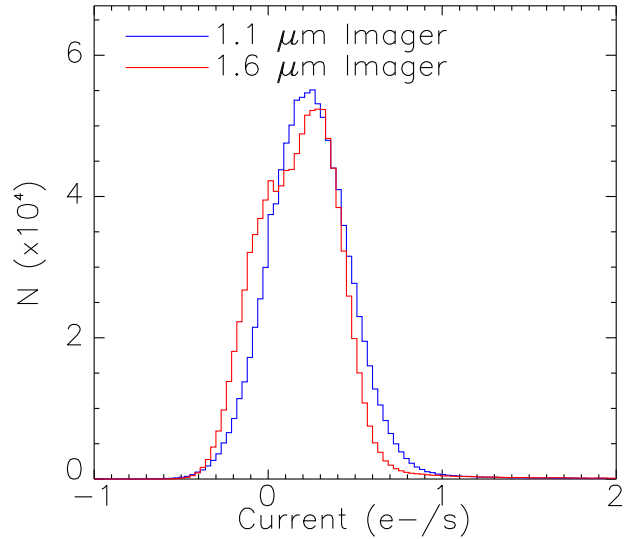,width=0.48\textwidth}
\caption{CIBER Imager dark currents for both cameras.  The mean dark
  current is $0.3 \,$\eps, which is consistent with the manufacturer's
  specifications for Hawaii-1 arrays operating near LN$_{2}$
  temperature.}
\label{fig:darkcurrent}
\end{figure}

The measured dark current also depends on the detector thermal
stability.  For the Imagers we require dark current stability of $0.1
\,$\eps, which is equivalent to $\pm 100 \, \mu$K/s given a
temperature coefficient of $1000 \,$e$^{-}$/K.  The Imager detector
arrays are controlled to $\pm 10 \, \mu$K/s both in the lab and in
flight, exceeding this specification \citep{Zemcov2012}.  In the
flight configuration with the cold shutter closed and the focal plane
under active thermal control, we achieve $\sim 0.3 \,$\eps\ mean dark
current, as shown in Figure \ref{fig:darkcurrent}.  The dark current
is measured frequently before launch as a monitor of the instrument
stability and is entirely consistent with the dark current measured in
the laboratory.  The stability of the dark current from run to run
indicates the dominant contributor to dark current is the array
itself, as opposed to temperature or bias drift.

\subsection{Noise Performance}
\label{sS:noise}

Measuring the REBL spatial power spectrum requires a precise
understanding of the noise properties of the array.  The array noise
introduces a bias that must be accounted and removed in
auto-correlation analysis, and determines the uncertainty in the
measured power spectrum.  The instrument sensitivity shown in Figure
\ref{fig:pwrspec} assumes the noise over the array is uncorrelated
between pixels.  Unfortunately, HgCdTe arrays exhibit correlated
noise, as described by \citet{Moseley2010}.  This noise is associated
with pickup from the clock drivers to the signal lines, with $1/f$ noise
in the multiplexer readout, and depending on the implementation, with
$1/f$ noise on the bias and reference voltages supplied to the array.

\subsubsection{Noise Model}

\begin{figure*}[ht]
\epsfig{file=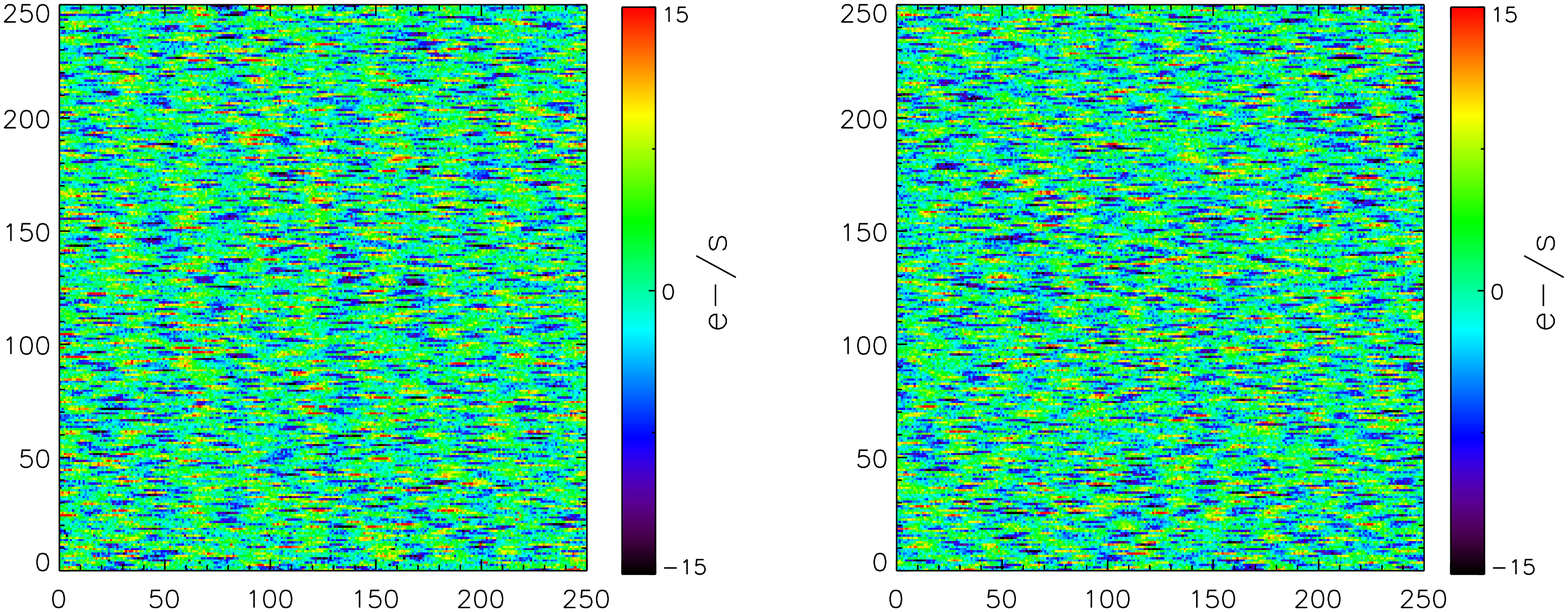,width=0.99\textwidth}
\epsfig{file=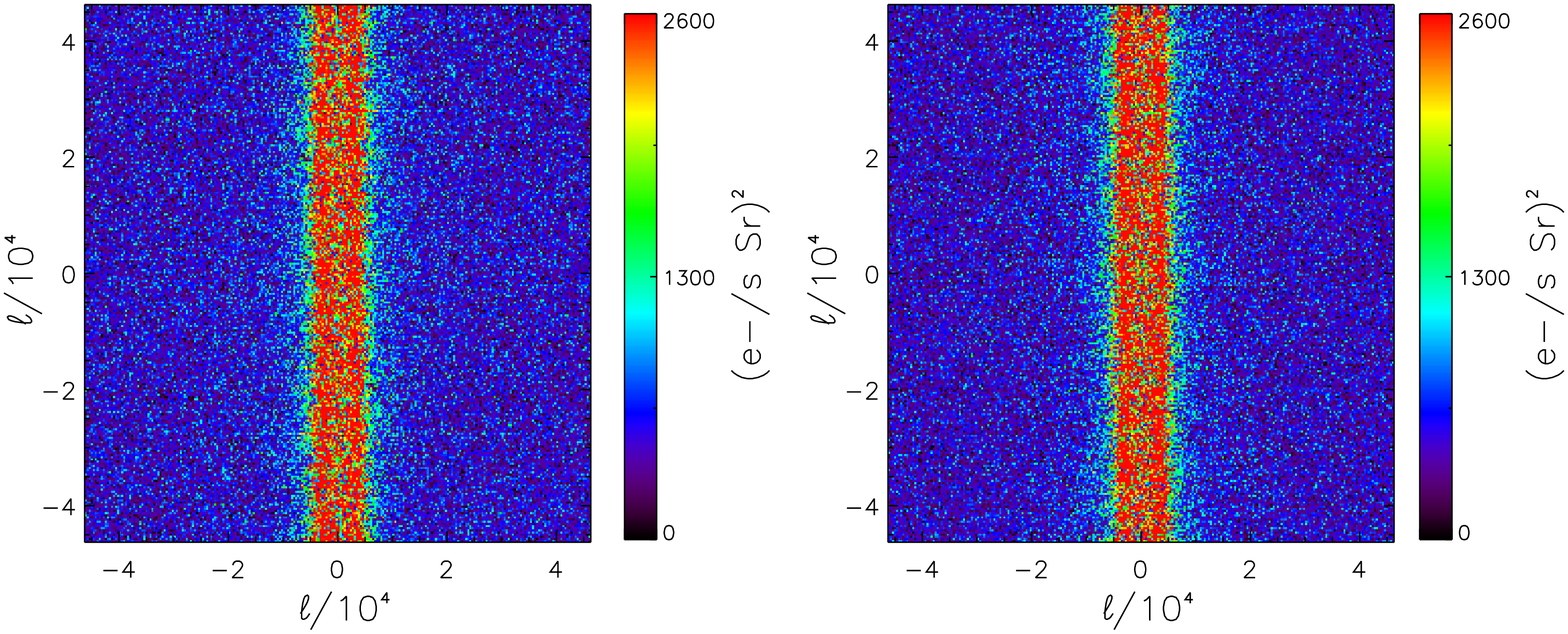,width=0.99\textwidth}
\caption{Images (top) and two dimensional power spectra (bottom) of the
  difference between two dark images, each obtained in a $50 \,$s integration.
  The upper and lower left hand panels show the image and power spectra of data
  taken minutes before flight, while the two right hand panels show the same for
  random realizations using the noise model presented in Section \ref{sS:noise}.
  The spatial scale of these images has been restricted to $250 \times 250$
  pixels to better show the spatial structure.  In both cases the read direction
  is horizontal along pixel rows.  The vertical structure in the two
  dimensional power spectra shows increased noise power in the read
  direction on scales $> 50$ pixels.  The noise model accurately captures
  this behavior, both in real and Fourier space.}
\label{fig:powerspectrum}
\end{figure*}

We characterized array noise using dark laboratory images and data
obtained just prior to flight.  We first took a series of dark
integrations to characterize the noise behavior similar to the $\sim
50\,$s integrations used in flight.  In the left hand panels of Figure
\ref{fig:powerspectrum} we show the two dimensional power spectrum of
the difference of two consecutive $50 \,$s laboratory integrations.
The spectrum shows enhanced noise at low spatial frequencies along the
read direction that is largely independent of the cross-read spatial
frequency, symptomatic of correlated noise in the readout.

\begin{figure}[ht]
\epsfig{file=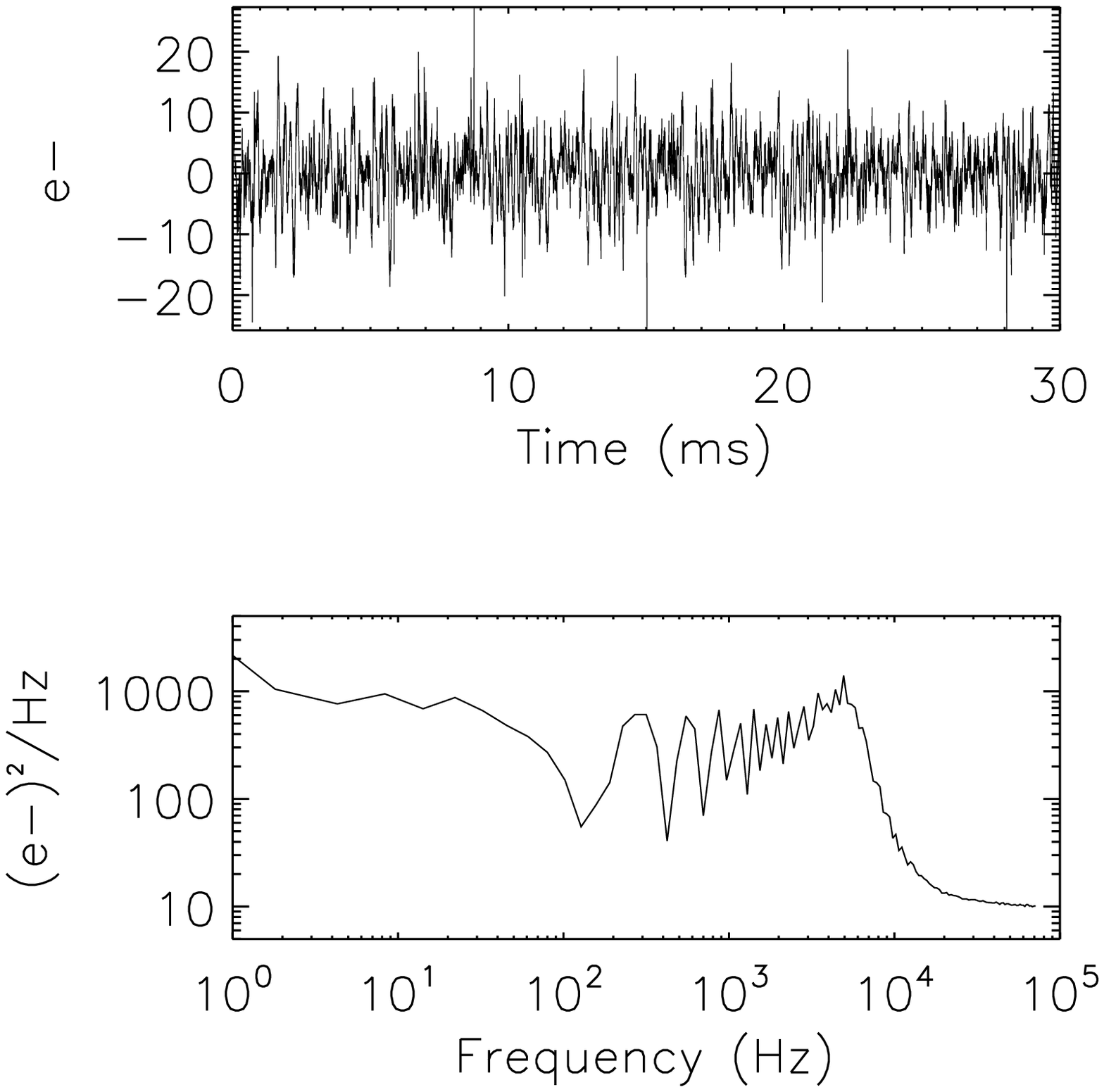,width=0.48\textwidth}
\caption{The upper panel shows $30 \,$ms of signal-subtracted time ordered
  data from the $1.6 \, \mu$m Imager.  The lower panel shows the noise
  spectrum derived from a longer such time series of reads over $50 \,$s.
  The noise increases at $\sim 10 \,$kHz, visible in the time stream in
  the upper panel as the characteristic scale of the noise at $\sim 0.5 \,$ms.
  The ringing visible in the power spectrum below $10 \,$kHz corresponds to
  the harmonics of the clock signals used to address the array.}
\label{fig:todpowerspectrum}
\end{figure}

We then generate an estimate of the noise by constructing time streams
for the array readout.  First, we determine the best fit slope and
offset for each pixel.  We then subtract this estimate of the photo
current signal in each pixel in each frame.  Finally, we form a
sequence of data for each of the four readout quadrants in the order
that the readout addresses individual pixels.  An example of
time-ordered data and its noise spectrum is shown in Figure
\ref{fig:todpowerspectrum}, exhibiting excess noise behavior similar
to that described in \citet{Moseley2010}.

The correlated noise in the readout may reduce the in-flight
sensitivity, and must be modeled to remove noise bias in the
auto-correlation power spectra.  While a full description of a noise
model of the flight data is outside the scope of this paper, we can
generate a model confined to the noise properties of the arrays
observed in laboratory testing.  This model is generated by producing
a Gaussian noise realization of the power spectrum given in Figure
\ref{fig:todpowerspectrum}.  This is used to generate random
realizations of time ordered data.  These data are mapped back into
raw frames, and fit to slopes and offsets to determine the images for
a full $50 \,$s integration.  To generate images like those shown in
Figure \ref{fig:powerspectrum}, we generate multiple images and
display the difference of two $50 \,$s images.  This formalism will be
extended to the flight data by adding photon shot noise from the
astrophysical sky, and correcting for source masking, in a future
publication.

\subsubsection{Estimated Flight Sensitivity}

\begin{figure*}[ht]
\epsfig{file=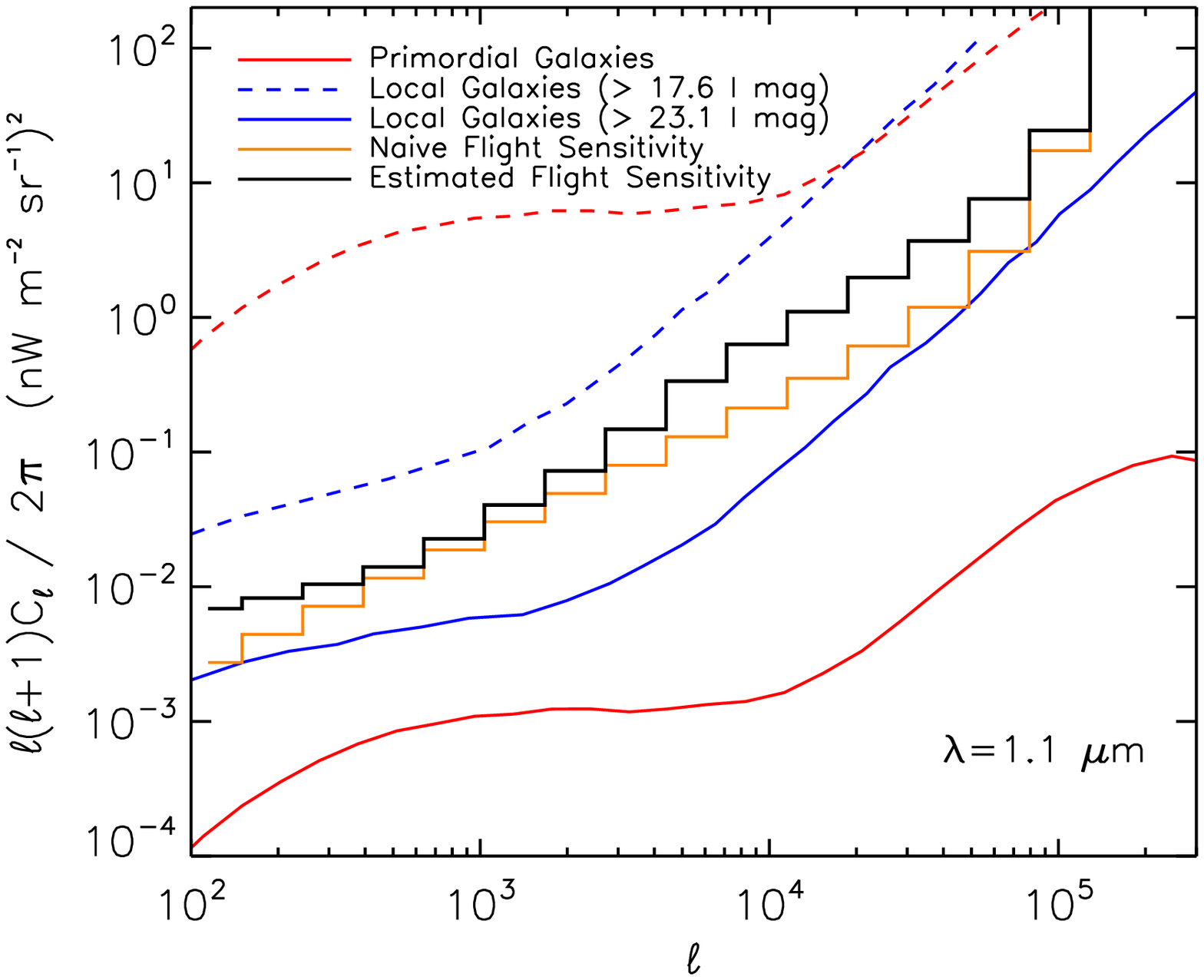,width=0.48\textwidth}
\epsfig{file=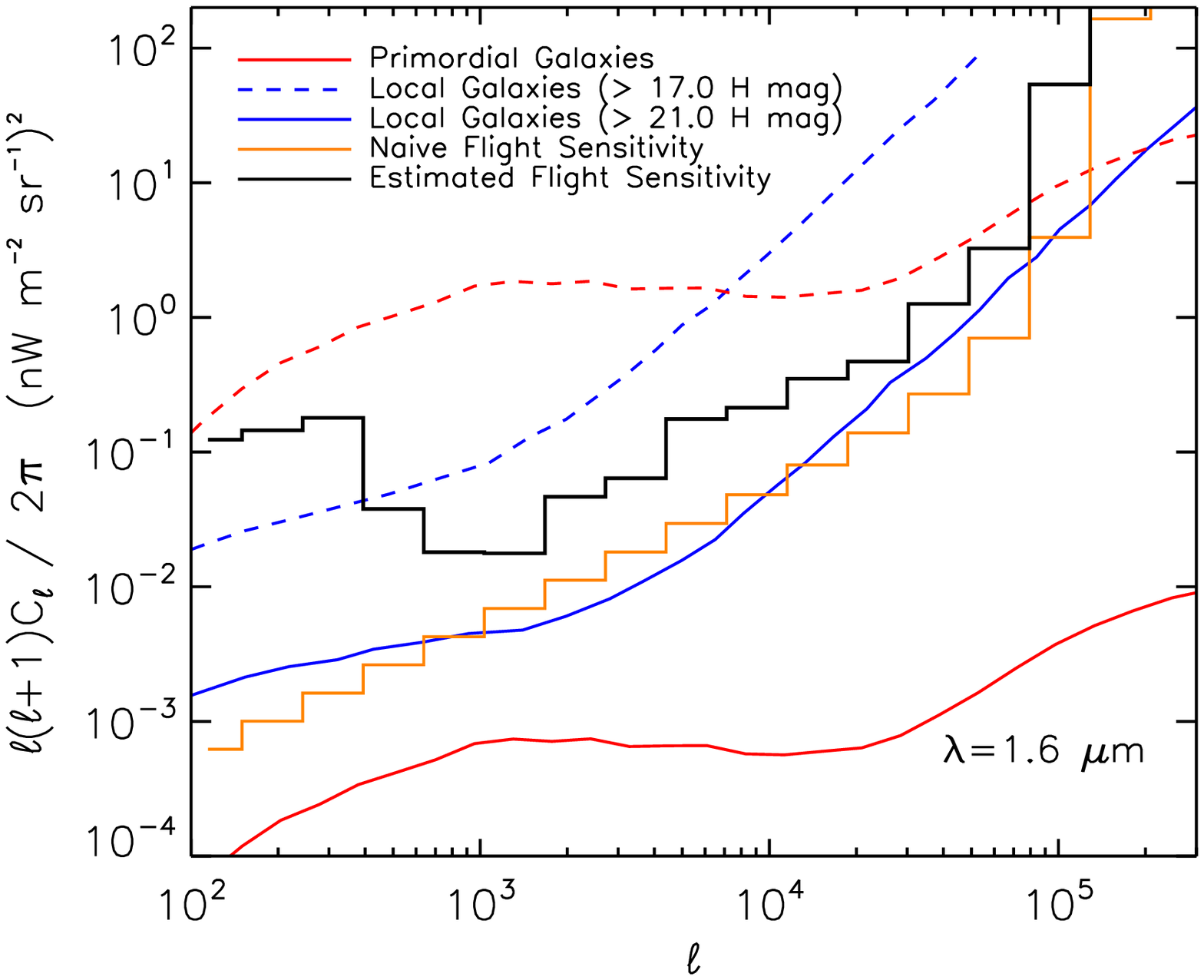,width=0.48\textwidth}
\caption{The Imager sensitivity to REBL fluctuations.  The left hand
  panel shows the estimated sensitivity for the $1.1 \, \mu$m channel,
  and the right for the $1.6 \, \mu$m channel.  In addition to the
  curves taken from Figure \ref{fig:pwrspec}, we show the sensitivity
  derived from laboratory data for both bands as described in the text
  using the same $\ell$ binning as the na\"{\i}ve sensitivity estimate
  shown by the orange curve.  The black curve is an estimate of the flight
  sensitivity, combining measured laboratory noise from an ensemble of $50 \,$s
  integrations, added with uncorrelated photon noise derived from the
  flight photo currents.  This estimate is for a single $50 \,$s integration,
  and does not include the effects of noise in the flat field or the
  loss of pixels from galaxy masking.}
\label{fig:newpssensitivity}
\end{figure*}

To calculate the effect of correlated noise on the final science
sensitivity, we take our sequence of dark laboratory images, calculate
the two dimensional power spectrum, and apply a two-dimensional
Fourier mask that removes modes sensitive to the excess low frequency
noise.  We remove these modes because they have a phase coherence in
real data that is not fully captured by the Gaussian noise model.
After Fourier masking, we calculate the spatial power in logarithmic
multipole bins.  We then evaluate the standard deviation in the
spatial power among eight dark images, and refer this to sky
brightness units using the measured calibration factors in Table
\ref{tab:imagersens}.  Because the laboratory data do not have
appreciable photon noise, we add an estimate of uncorrelated
photon noise from the flight photo currents.  We compare this empirical
determination of the noise with the na\"{i}ve sensitivity calculation
in Figure \ref{fig:newpssensitivity} \citep{Knox1995}.  The empirical
noise is close to the na\"{i}ve calculation on small spatial
frequencies, but is degraded by correlated noise on large spatial
scales.  However the instrument is still sufficiently sensitive to
easily detect the optimistic REBL power spectrum.  For future experiments,
one may address the reference pixels in Hawaii-RG arrays to mitigate
the effects of correlated noise.

\subsection{Detector Non-linearity and Saturation}
\label{sS:effects}

\begin{figure}[ht]
\epsfig{file=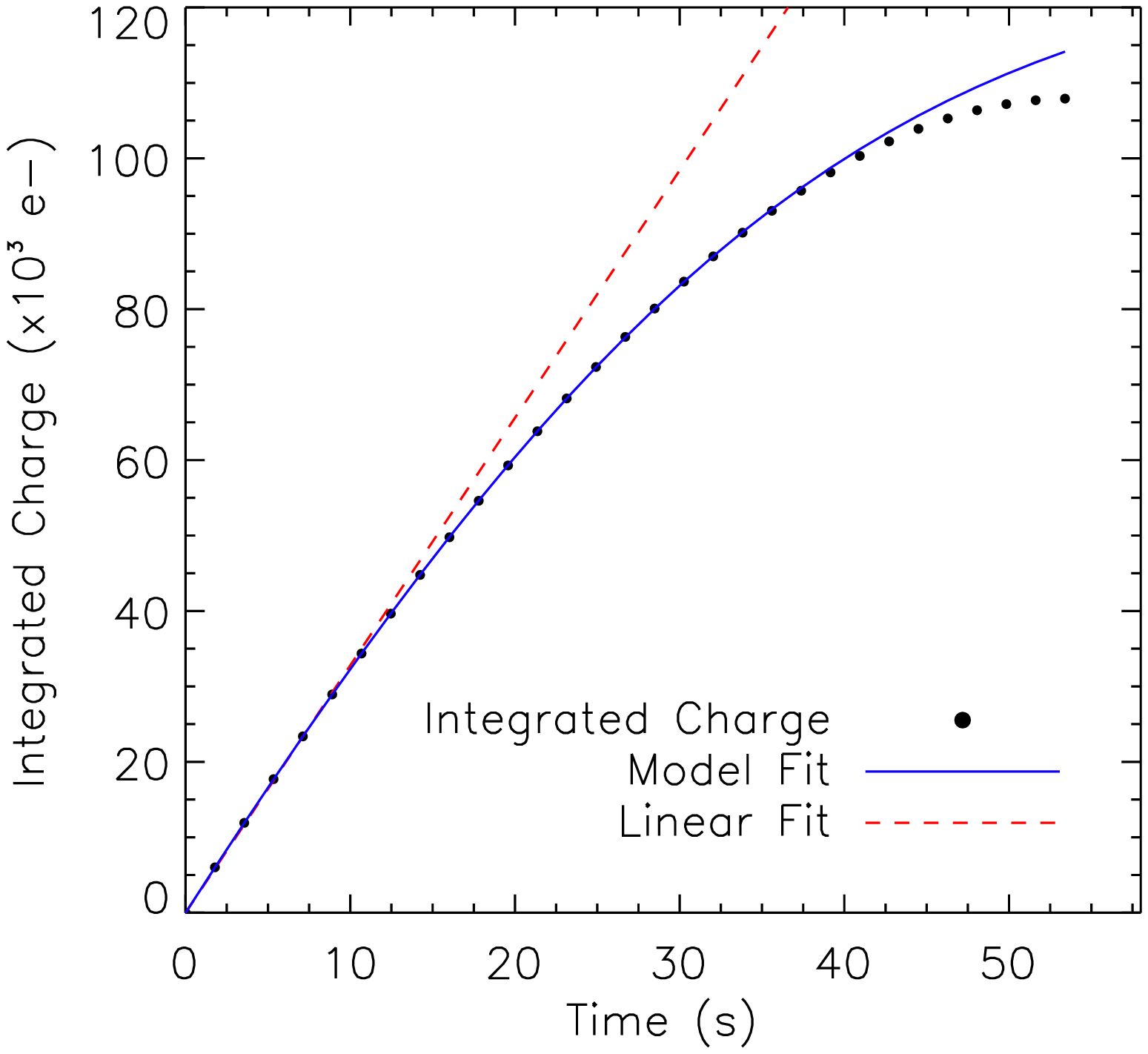,width=0.48\textwidth}
\caption{Integrated signal as a function of time for a typical Imager
  pixel. The black data show subsequent reads of the Imager detector
  for an incident brightness of $\sim 3500 \,$\eps.  The dashed red
  line shows the linear model matching the slope of the first $10 \,$s
  of the integration.  Finally, the blue line is a fit to the model
  from \cite{Bies2011}, which agrees well with the data.}
\label{fig:linearity}
\end{figure}

The Imager detectors have a dynamic range over which the response
tracks the source brightness in a linear fashion.  As is typical for
Hawaii-1 detectors, the full well depth is measured to be $\sim 10^{5}
\,$e$^{-}$; however, the detectors begin to deviate from linearity
well before this.  In order to flag detector non-linearity, we find
pixels with different illumination levels and track their behavior
during an integration.  Figure \ref{fig:linearity} shows the typical response
of a pixel to a bright $\sim 3500 \,$\eps\ source over time.  This plot shows
a deviation from the linear model which is large at half the full well
depth.  Except for a few bright stars, Imager flight data are well within
the linear regime.  Pixels with an integrated charge greater than $7000 \, e^{-}$
have a non-linearity $\sim 1 \%$ are simply flagged and removed from further
analysis, amounting to a pixel loss of $< 0.5 \%$ over the array.

\subsection{Focus and Point Spread Function}
\label{sS:focus}

CIBER is focused in the laboratory by viewing an external collimated
source through a vacuum window.  Early on in focus testing we found
that the best focus position depended on the temperature of the
optics.  Thermal radiation incident on the cameras can heat the front
of the optics and affect their optical performance due to both
differential thermal expansion and the temperature-dependent refractive
index of the lenses.  We reduced the incident thermal radiation
by installing two fused silica windows in front of the cameras for
laboratory testing.  The cold windows themselves are $125 \,$mm
diameter, $5 \,$mm thick SiO$_{2}$, operating at a temperature of $120
\,$K, and have $1/10$ surface flatness and $< 5"$ wedge.  As described
in \citet{Zemcov2012}, these windows are thermally connected to the
radiation shield to direct the absorbed thermal power to the
liquid nitrogen tank instead of routing the power through the optical
bench where it would produce a temperature gradient across the optics.

\begin{figure}[ht]
\epsfig{file=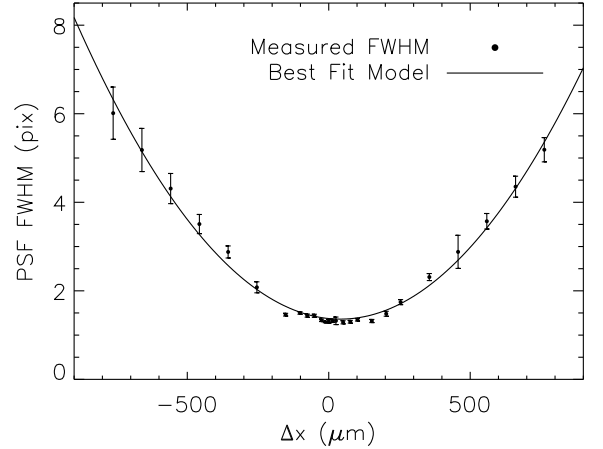,width=0.48\textwidth}
\caption{The variation of the PSF width measured in the laboratory as
  a function of collimator focus position $\Delta x$ shifted away from
  its best focus position.  At each collimator position we measure the
  PSF by fitting a Gaussian and determining its full width at half
  maximum (FWHM) and uncertainty.  The points show the data and the
  black line the best fit parabola to the points, yielding the best
  estimate of the focus position of the Imager instrument.  The curve
  is consistent with the $f/4.95$ focal ratio, where the array pixels
  are $18 \times 18 \, \mu$m and subtend $7 \times 7$ arcseconds on
  the sky.}
\label{fig:labpsf}
\end{figure}

With the cold windows in place, we measure focus using a collimator
consisting of an off-axis reflecting telescope with a focal length of
$1900 \,$mm, a $235 \,$mm unobstructed aperture, and an $8 \, \mu$m pinhole
placed at prime focus. Since the focus position of the instruments is fixed, we
scan the pinhole through the focus position of the collimator to find the displacement
from collimator best focus at which each Imager has its best focus.  This
procedure is repeated at the center of the array, the corner of each
quadrant, and in the center again as a check of consistency.  Figure
\ref{fig:labpsf} shows data from such a test.  If the focal plane focal
distance is found to be outside the $\pm 80 \, \mu$m focal depth of the
Imagers, we mechanically shim the focal plane assembly to the best focus
position and remeasure the focus.  We verify the focus position
before and after pre-flight vibration testing, performed for each flight,
to ensure that the focus will not change in flight.

We measure the point spread function (PSF) in flight using stars as
point sources.  Given the large number of sources detected in each
field, a measurement of the average PSF across the array can be
obtained by fitting all of the bright sources. In fact, because the
astrometric solution of the images allows us to determine source
positions more accurately than a single pixel, and because the pixels
undersample the PSF of the optics, stacking sources gives a more
accurate determination of the central PSF.  To generate the stack, the
region containing each source is re-gridded to be $3 \times$ finer than
the native resolution.  The finer resolution image is not interpolated
from the native image, rather, the nine pixels which correspond to a
single native pixel all take on the same value.  However, when we stack
the re-gridded point source images we center each image based on the known
source positions, and thus the stacked PSF is improved using this sub-pixel
prior information.

To measure the extended PSF, we combined data from bright sources, which
saturate the PSF core, with faint sources that accurately measure the PSF
core.  We generate the core PSF by stacking sources between 16.0 and 16.1 Vega
magnitudes from the 2MASS catalog \citep{Skrutskie2006}, which provides
a set of sources that are safely in the linear regime of the detector.  The
source population is a combination of stars and galaxies, however with $7''$
pixels, galaxies are unresolved.  As a check, this same analysis was repeated
for sources between 15.0 and 15.1, and 17.0 and 17.1 Vega magnitudes.  The PSF
generated from these magnitude bands agreed with the nominal PSF.

To measure the extended PSF, we stack bright sources between 7 and 9 Vega
magnitudes from the 2MASS catalog.  Since these bright sources are heavily
saturated, the best fit Gaussian is only fit to the outer wings for
normalization.  After the core and extended PSFs are created, we find they
agree well in the region between $r \sim 13 \,$arcsec, inside of which the
bright sources are saturated, to $r \sim 30\,$arcsec, where the faint sources
are limited by noise.

\begin{figure}[htb]
\epsfig{file=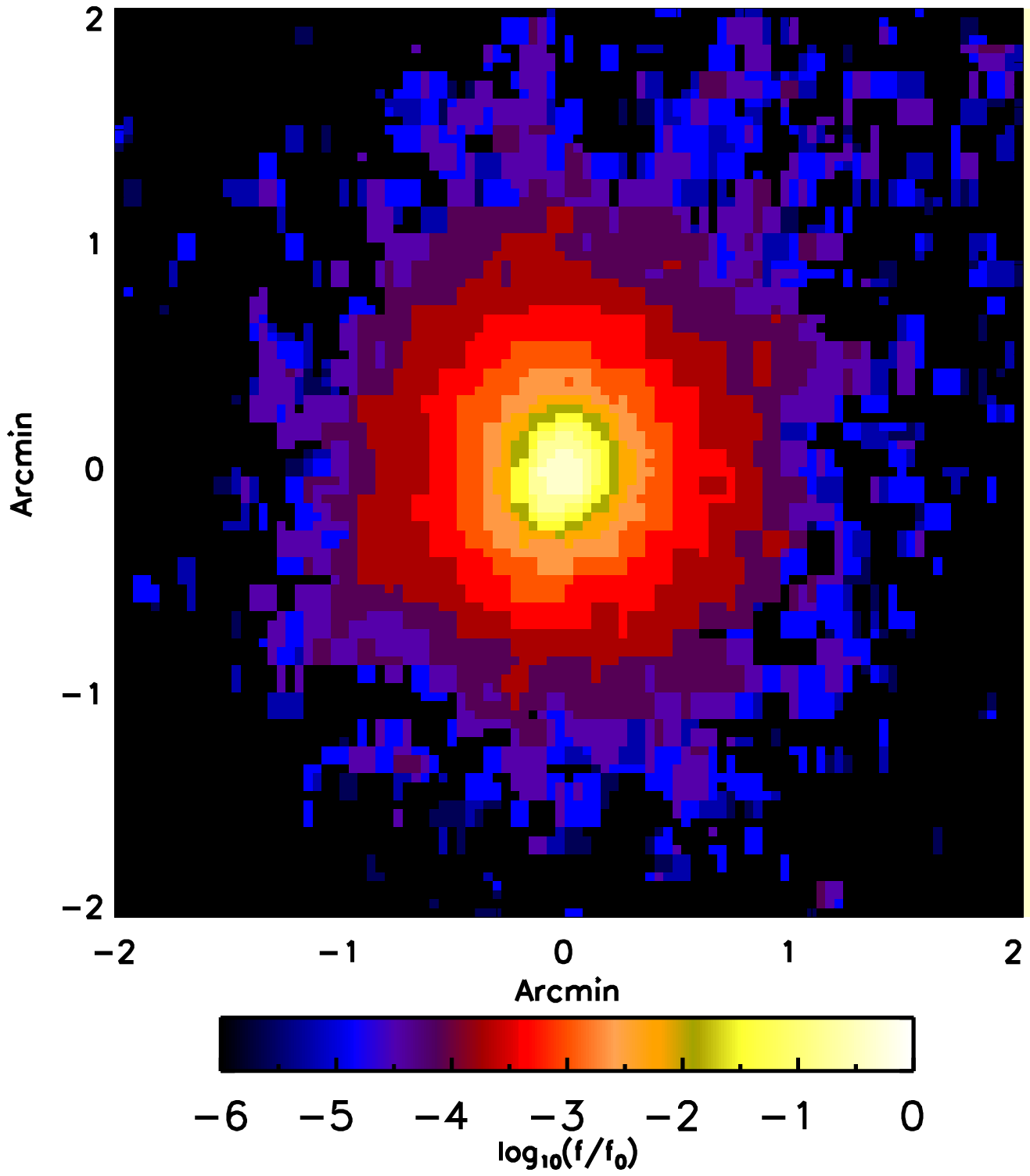,width=0.57\textwidth}
\epsfig{file=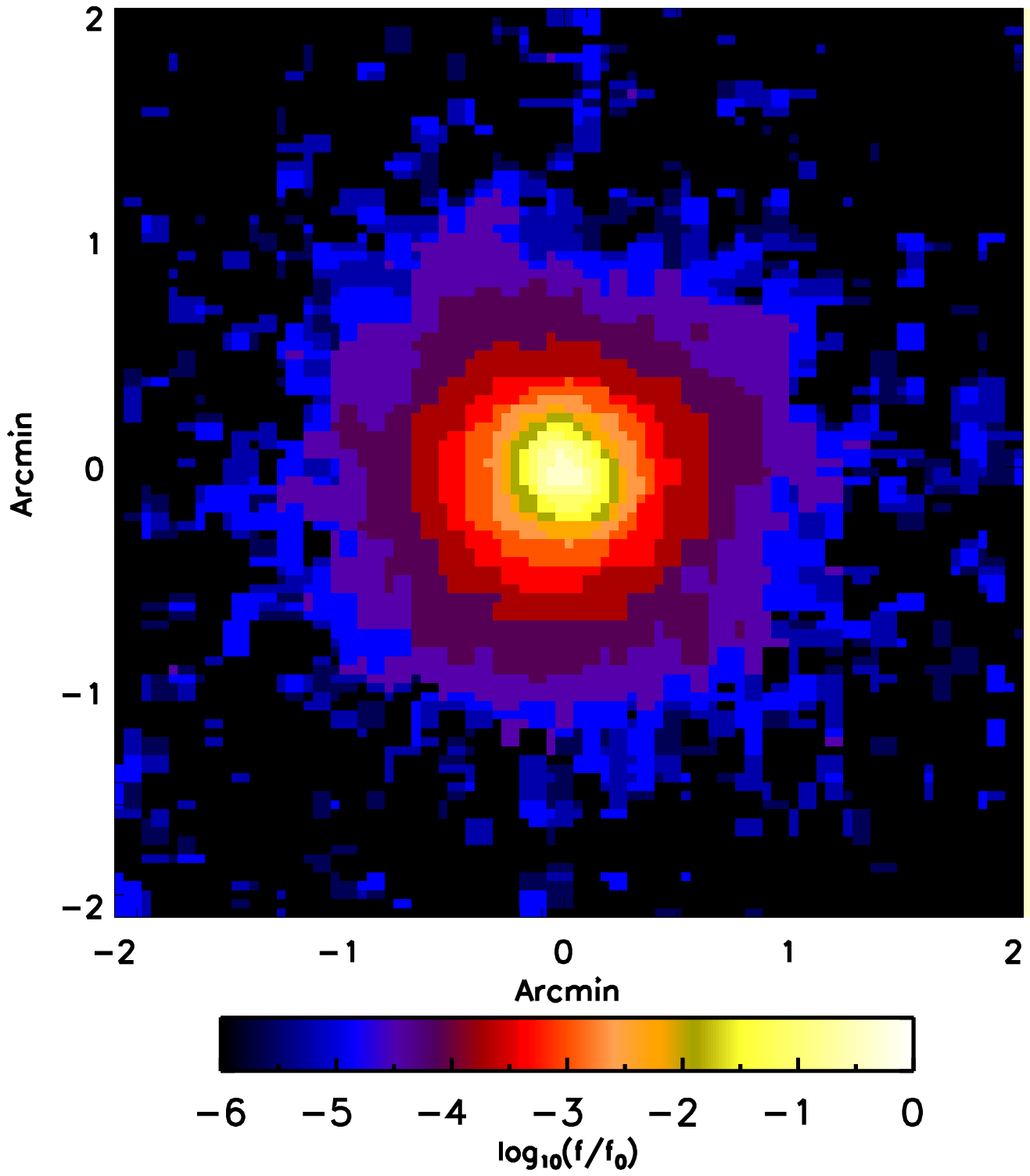,width=0.57\textwidth}
\caption{The $1.1 \, \mu$m (left) and $1.6 \, \mu$m (right) Imager PSFs measured
  using stacked flight images from a combination of bright and faint sources as
  described in the text.  The Imager PSF has a bright core with a faint
  extension to $r \sim 1'$, and is circularly symmetric.}
\label{fig:flightpsf}
\end{figure}

\begin{figure}[htb]
\epsfig{file=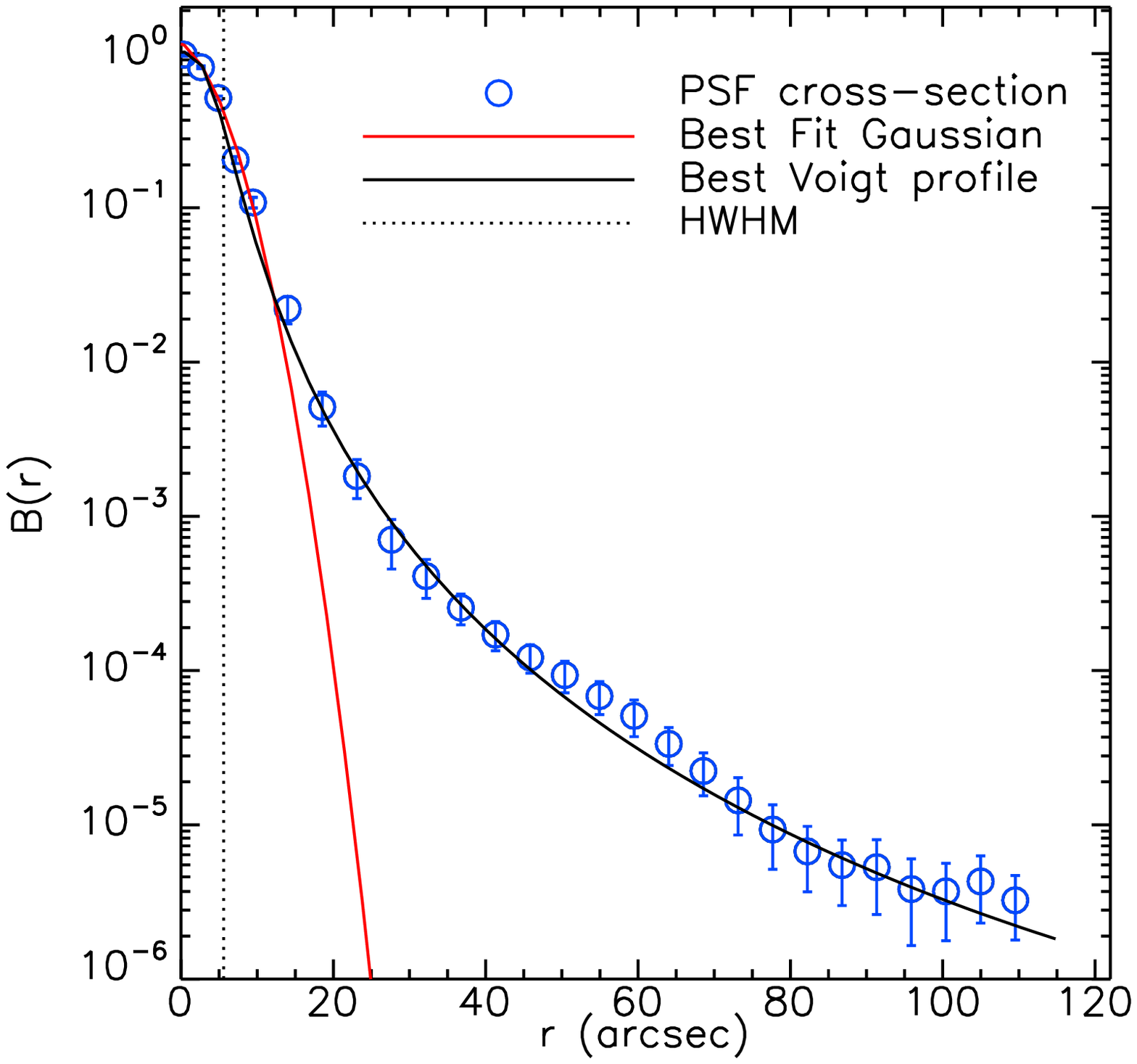,width=0.48\textwidth}
\epsfig{file=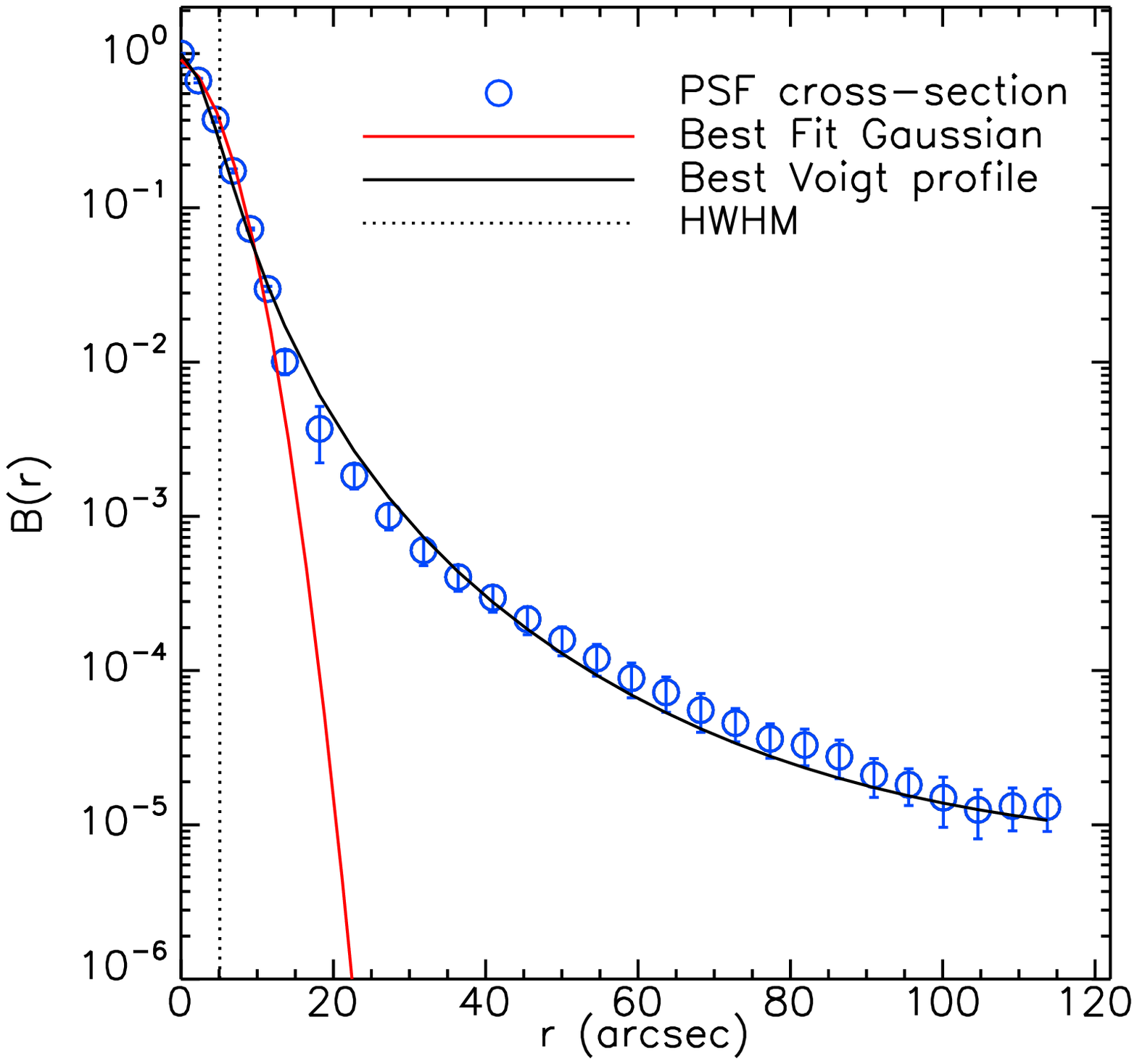,width=0.48\textwidth}
\caption{The radial profile of the $1.1 \, \mu$m (left) and $1.6 \, \mu$m
  (right) Imager flight PSFs from Figure \ref{fig:flightpsf} (blue circles).
  The red curve shows the best fit Gaussian to the PSF core, while the black
  curve shows a best fit Voigt (i.e.~the convolution of a Gaussian and
  Lorentzian) function to the extended PSF.  This is indicative of scattering
  in the optical components.  Finally, the black dash-dotted line shows
  the HWHM of the PSF, which matches the value measured in the
  laboratory.}
\label{fig:ringpsf}
\end{figure}

We synthesize the full PSF by matching the amplitudes of the
core and extended PSFs in the overlap region, producing the smooth two
dimensional PSF shown in Figure~\ref{fig:flightpsf}.  The radial
average of this full PSF is shown in Figure \ref{fig:ringpsf} and
highlights that the core PSF is consistent with the laboratory focus data.
However, the extended PSF deviates significantly from this approximation
and is better described by a Voigt profile shape, characteristic of
scattering in the optical components.

The extended PSF is essential for determining the appropriate mask to
apply for bright sources.  The diameter of the PSF mask is adjusted
based on the brightness of the source, and pixels above a given flux
are cut.  The cut is calculated by simulating all sources in either
the 2MASS or \textit{Spitzer}-NDWFS catalogs using their known fluxes
and the Imager PSF.  The cut mask is generated by finding all points
on this simulation with fluxes $> 3.3 \,$\nw\ and $1.8 \,$\nw\ at
$1.1$ and $1.6 \, \mu$m, respectively.  This masking algorithm retains
$\sim 50 \,$\% of the pixels for a cutoff of $18$ Vega mag, and
$\sim$30\% of the pixels for a cutoff of $20$ Vega mag.  To test the
cutoff threshold, we simulate an image of stars and galaxies and find
that, cutting to $20$ mag, the residual spatial power from masked
sources is $< 8 \times 10^{-2} \, nW^{2} m^{-4} sr^{-2}$ at $\ell
= 10^{4}$, comparable to the instrument sensitivity shown in Figure
\ref{fig:newpssensitivity}.

\subsection{Off-axis response}
\label{sS:offaxis}

The Imagers must have negligible response to bright off-axis sources,
including the ambient-temperature rocket skin and shutter door, and
the Earth.  As described in \citet{Zemcov2012}, we added an extendable
baffle to eliminate thermal emission from the rocket skin and
experiment door, heated during ascent by air friction, from illuminating
the inside of the Imager baffle tube and scattering to the focal plane.

\begin{figure}[htb]
\epsfig{file=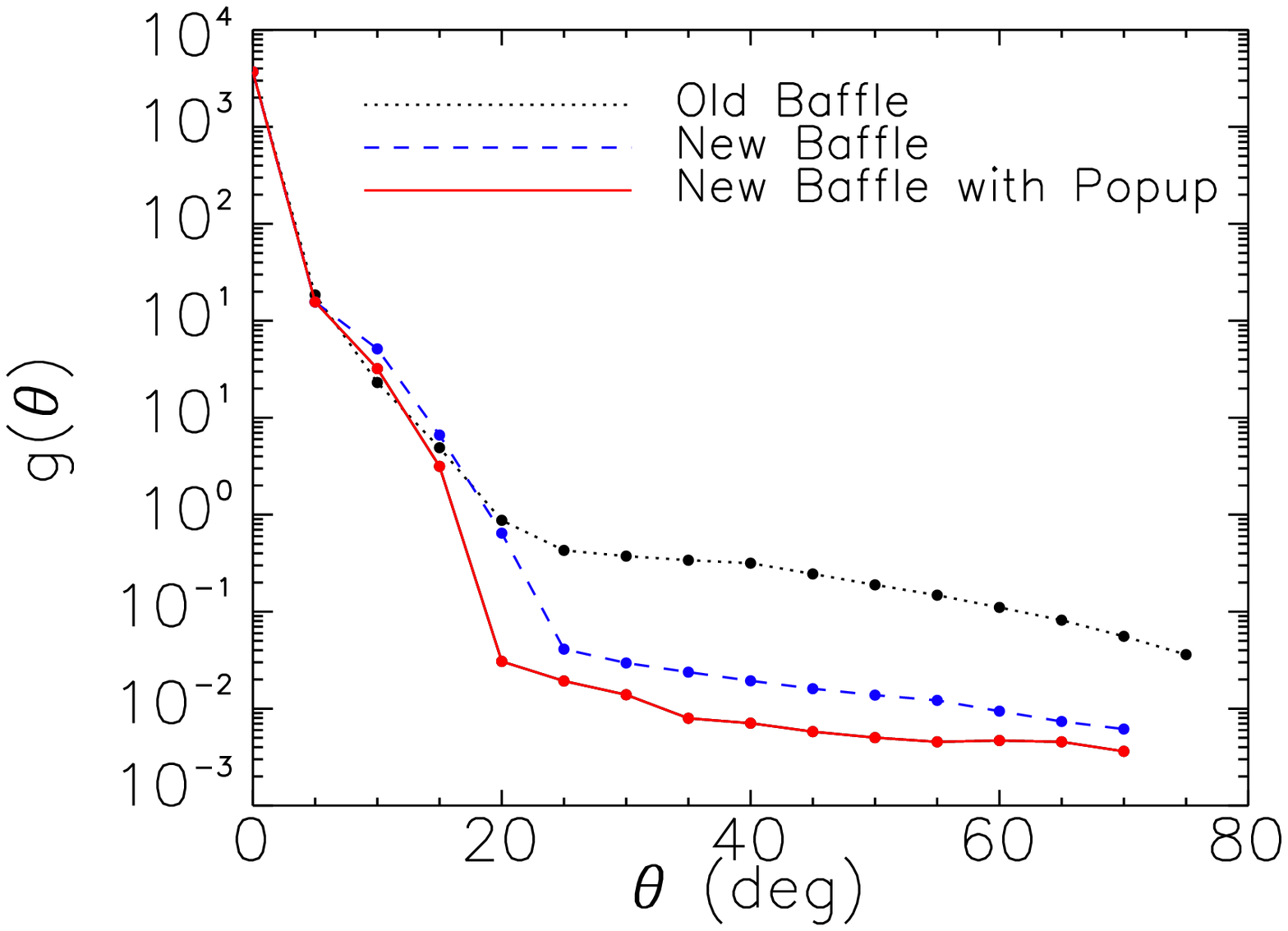,width=0.48\textwidth}
\caption{The Imager telescope gain function, measured with the
  anodized fixed black baffle tube used in the first flight (dotted
  black line), an improved fixed baffle tube with a better laser black
  optical coating (Epner Technology Inc., dashed blue line), and the
  combination of the improved fixed baffle with an extendable baffle
  used in the second flight (solid red line).  Details of the optical
  baffling can be found in \citet{Zemcov2012}.}
\label{fig:offaxis}
\end{figure}

We measured the off-axis response of the full baffle system following
the methodology in \citet{Bock1995}.  We replaced the Hawaii-1 focal
plane array with a single optical photo diode\footnote{Hamamatsu Si
$10 \times 10\,$mm$^{2}$ detector part number S10043.}  detector and
measured the response to a distant chopped source (see
\citealt{Tsumura2012} for a complete treatment of the measurement).
The telescope gain function,
\begin{equation}
\label{eq:gth}
g(\theta) = \frac{4 \pi}{\Omega} G(\theta) ,
\end{equation}
where $\Omega$ is the solid angle of the detector and $G(\theta)$ is
the normalized response to a point source is the quantity of interest
for immunity to off-axis sources in surface brightness measurements
(\citealt{Page2003}) and is independent of the optical field of view.
The gain function was measured for three baffle configurations and is
shown in Figure \ref{fig:offaxis}.  The improvement from blackening
the baffle tube and adding an extendable baffle section is notable for
angles $\theta > 20^{\circ}$.  The stray light level from the Earth
is given by
\begin{equation}
\label{eq:Istray}
I_{\mathrm{stray}} = \frac{1}{4 \pi}\int g(\theta)
I_{\earth}(\theta,\phi)d \Omega ,
\end{equation}
where $I_{\earth}$ is the surface brightness of the Earth, and
$I_{\mathrm{stray}}$ is the apparent surface brightness of stray light
referred to the sky.  Following the calculation described in
\citet{Tsumura2012}, we estimate that during the second flight CIBER
observations of the fields listed in Table \ref{tab:ancfields}, where the
Earth's limb is $> 72^{\circ}$ off-axis, the stray light level is
calculated to be $2 \,$\nw\ and $1 \,$\nw\ in the $1.6$ and $1.1 \, \mu$m
channels, respectively.

This level of stray light is quite small but not completely
negligible, and potentially problematic in an anisotropy measurement
depending on its morphology over the field of view.  To quantify how
stray light affects our measurements, we calculated the spatial power
spectrum of the difference between two images, Bo\"{o}tesA -
Bo\"{o}tesB which are separated by only $2^{\circ}$ on the sky and
taken at nearly the same Earth limb avoidance angle, and Bo\"{o}tesA -
NEP, from second flight data (see section \ref{S:performance}).  We
find that the power spectra of these differences are the same to
within statistical noise, and that the spatial fluctuations of the
stray light signal are negligible.

We plan to observe these fields again in future flights at different
Earth limb avoidance angles, including angles greater than
$90^{\circ}$.  The cross-correlation of such images from different
flights is highly immune to residual stray light.

\subsection{Flat Field Response}

The instrumental flat field, which is the relative response of each
detector pixel to a uniform illumination at the telescope aperture, is
determined in flight by averaging observations of independent fields.
Additionally, the flat field can be independently measured in the laboratory
before and after flight as a check for systematic error.  The laboratory flat
field response is measured by illuminating the full aperture of a camera with
the output of an integrating sphere.  The sphere is illuminated with a
quartz-tungsten halogen lamp which is filtered to produce an approximately solar
spectrum at the output of the sphere, mimicking the spectrum of ZL.

The sphere was measured by the manufacturer to have uniformity as a function of
angle to better than $5 \times 10^{-3}$ over $10^{\circ} \times 10^{\circ}$.  We
scanned a small collimating telescope with a single pixel over the aperture, and
determined that the sphere has angular uniformity to better than $1 \times 10^{-3}$
over the $2^{\circ} \times 2^{\circ}$ Imager field of view.  We also measured the
spatial uniformity over the output port and saw no evidence of non-uniformity to
$< 7 \times 10^{-3}$ over an 11 cm aperture.

\begin{figure*}[ht]
\epsfig{file=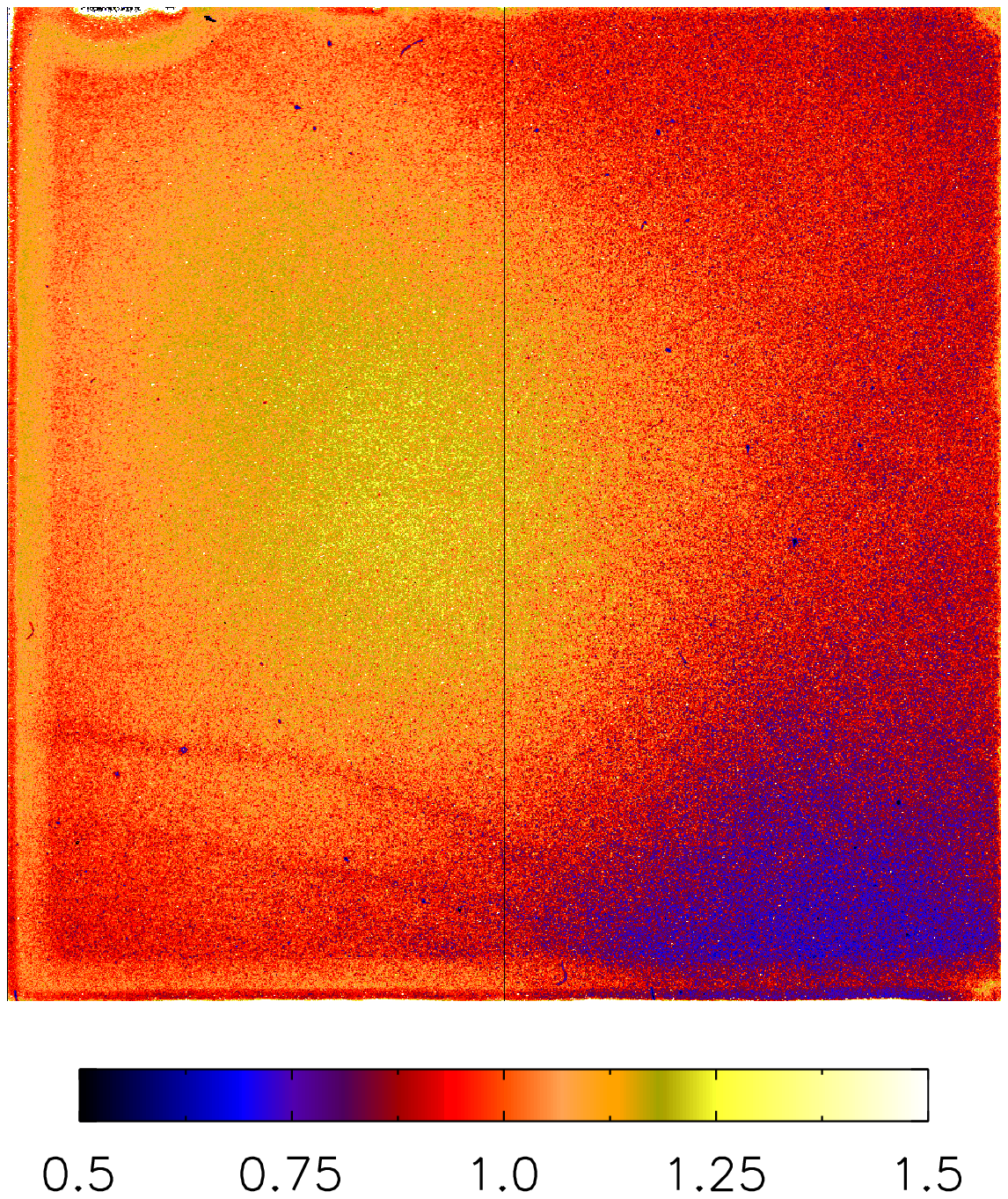,width=0.48\textwidth}
\epsfig{file=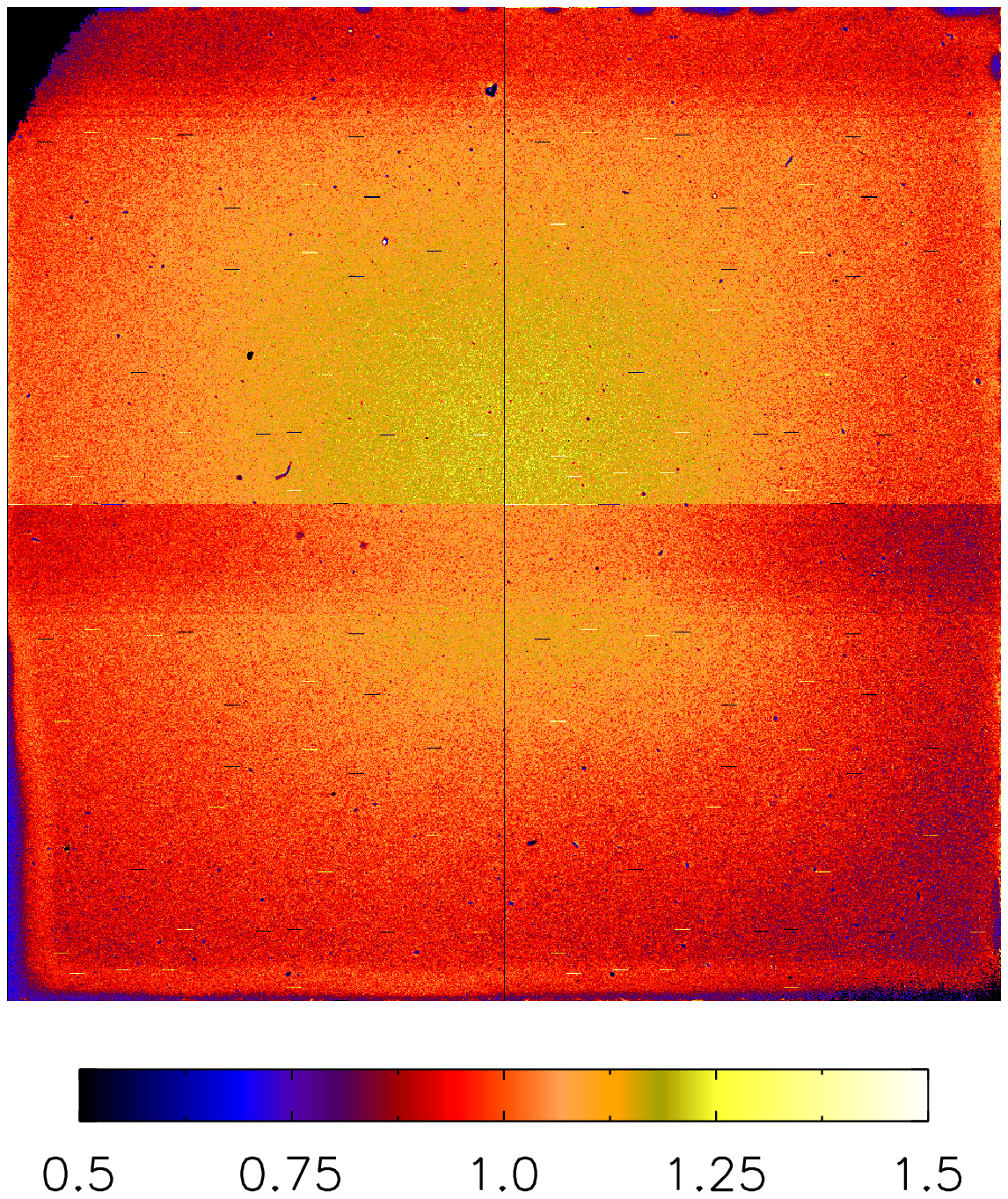,width=0.48\textwidth}
\caption{The $1.1 \, \mu$m and $1.6 \, \mu$m Imager flat fields as
  measured in the lab using the apparatus described in
  \citet{Zemcov2012}.  The average response has been scaled to $1.0$
  in this image, which shows the typical relative responsivity
  performance of the Hawaii-1 arrays in conjunction with the optics.
  The RMS variation in the pixel responsivities is $0.09$ at $1.1 \,
  \mu$m and $0.12$ at $1.6 \, \mu$m.}
\label{fig:flatfield}
\end{figure*}

To eliminate any effects from vacuum and thermal windows, we house the integrating
sphere inside a vacuum chamber which mates to the front of the cryostat in place
of the shutter door (see \citet{Zemcov2012} for details).  Light is fed into the
sphere from outside of the vacuum box so that the lamp can be chopped at the
source, allowing us to remove the thermal background.  An example flat field
measurement for the $1.1 \, \mu$m camera is shown in Figure \ref{fig:flatfield}.

The laboratory data are fitted over a limited period of the integration following
array reset so as to avoid an appreciable error from non-linearity, as described in
section \ref{sS:effects}, taking into account the minimum well depth of all pixels
in the array.  The instruments have a residual response to thermal infrared radiation
in the laboratory with a typical photo current of $600 \, e^{-}/s$ in the $1.6 \, \mu$m
array, which therefore limits the linear integration period to $\sim 5$ s.  We obtained
interleaved data with the source on and off to monitor and subtract this thermal
background.  After accounting for these effects, the final statistical accuracy of
the laboratory flat field images shown in Figure~\ref{fig:flatfield} is $1.6 \%$
per pixel.  Laboratory flat fields were measured before and after the second flight
to quantify the reproducibility of the lab flat field response.  We binned
$1.6 \, \mu$m camera laboratory flat field images into 64 $(15 \times 15)$
arcminute square patches in order to reduce statistical noise, and found the binned images
agree to $< 1 \% (1 \, \sigma$).  The agreement between the flight and laboratory flat fields
requires a full reduction of the flight data and will be presented in a future science paper.

\section{Modifications Following the First Flight}
\label{S:mods}

The Imagers were flown on the CIBER instrument on a Terrier Black
Brant sounding rocket flight from White Sands Missile Range in
2009 February.  Many aspects of the experiment worked well, including
the focus, arrays and readout electronics, shutters, and calibration
lamps.  However, we also found several anomalies that led to
modifications for subsequent flights.

\subsection{Thermal Emission from the Rocket Skin}
\label{sS:thermalemission}

The instruments showed an elevated photon level during the flight due
to thermal emission from the rocket skin, heated by air friction during
ascent, scattering into the optics.  The edge of the skin near the shutter
door can directly view the first optic and the inside of the static baffle.
This thermal response was pronounced at long wavelengths, as traced by the LRS
\citep{Tsumura2010}.  The $1.6 \, \mu$m Imager was more affected by thermal
emission than the $1.1 \, \mu$m Imager, as expected from its longer wavelength
response, giving 40 and 7 times the predicted photo current, respectively.

The measured thermal spectrum with the LRS should not produce a significant
photo-current in the $1.1 \, \mu$m Imager, as the band is supposed to cut off
at $1.32 \, \mu$m.  The excess photo-current indicates the $1.1 \, \mu$m Imager
has some long wavelength response.  The array response may continue somewhat
beyond $2.5 \, \mu$m, as the band-defining filters provided blocking out
to just $2.5 \, \mu$m and then open up.  Also as with the NBS \citep{Korngut2012}
the filters may not attenuate scattered light at large incident angles
as effectively as at normal incidence.  The brightness observed by the
$1.6 \, \mu$m Imager is 6 times higher than the band-averaged LRS brightness.
This could be due to a combination of the higher stray light response in
the $1.6 \, \mu$m Imager, and the filter blocking issues mentioned above.
We installed an additional blocking filter providing $< 0.1 \%$ transmittance
from $2.4 \, \mu$m to $3.0 \, \mu$m for both imagers.

We modified the front of the experiment section to better control the
thermal and radiative environment at the telescope apertures.  Most
notably, we added extendable baffles to each of the instruments to
eliminate all lines of sight from the skin to the optics or the inside
surfaces of the baffle tubes.  \citet{Zemcov2012} details the design
of these baffles and the other changes made to the experiment section
front end.  Thermal emission is not detectable in the Imagers in the
second flight, and is at least 100 times smaller than the first flight
in the LRS data.

\subsection{Rings and Ghosts from Bright Sources}
\label{sS:imagerrings}

\begin{figure}[htb]
\epsfig{file=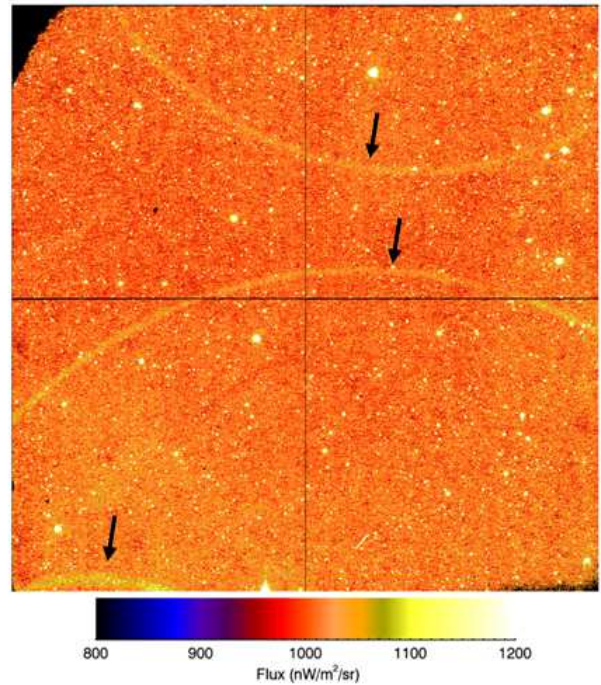,width=0.48\textwidth}
\caption{$1.1 \, \mu$m image of the Bo\"{o}tes A field from CIBER's
  first flight showing rings which were later traced to reflections
  off components inside the Imagers, namely the lens mounts and
  instrument walls.  As a guide the brightest rings are indicated with
  arrows.  There are three separate populations of reflections which
  produce these rings.  All sources which fall into their angular
  response regions will produce a ring, though only sources brighter
  than magnitude $\sim 4$ produce rings which are visible by eye.
  These rings produce excess power in the science power spectrum, but
  were eliminated by modifying the optics for the second flight.}
\label{fig:bootesrings}
\end{figure}

\begin{figure}[htb]
\epsfig{file=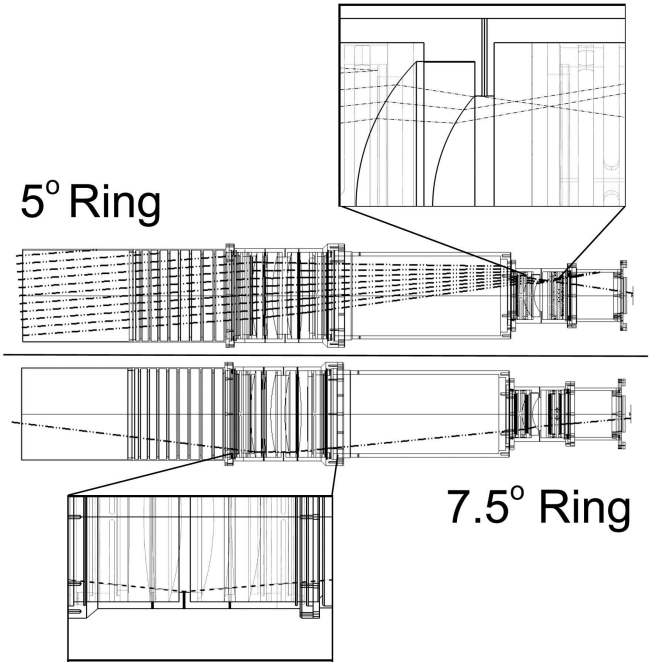,width=0.48\textwidth}
\caption{Ray trace from an off-axis source which produces the rings
  observed at the focal plane.  The first class of rings (labeled as
  $5^{\circ}$ in the Figure) are caused by glancing reflections off a
  flange supporting the back lens.  The second class of rings (labeled
  as $7.5^{\circ}$) is produced by glancing reflections off flanges
  and lens holders in the front set of optics.  For the second flight,
  these surfaces were cut back and grooved to reduce the glancing
  reflectance, removing the rings to a negligible level, as verified
  by laboratory measurements.}
\label{fig:ringraytrace1}
\end{figure}

During analysis of the first flight data, we discovered that bright
objects outside of the Imager field of view create diffuse rings in
the final images, as shown in Figure \ref{fig:bootesrings}.  Upon
further analysis, we found that each of these rings was centered on a
bright star outside the geometric field of view.  The rings were
caused by reflections off internal elements of the telescope assembly,
as illustrated in Figure \ref{fig:ringraytrace1}.  There are two
general classes of rings in the first flight images, though the second
class contains two distinct populations; we denote these ring
populations 1, 2 and 3 below.  Table \ref{tab:rings} gives details of
the ring populations including their angular extent and coupling
coefficients.

\begin{table*}[htb]
\centering
\caption{First Flight Imager Ring Parameters.}
\begin{tabular}{lccccc}
\hline
Ring Type & $\theta_{\mathrm{min}}$ & $\theta_{\mathrm{max}}$ &
\multicolumn{3}{c}{$\int d \phi I_{\mathrm{ring}}(\phi) / \int I_{0}$} \\

& & & Pre-fix & Post-fix ($3 \sigma$) & Reduction in $C_{\ell}$ \\ \hline
$1.1 \mu$m Imager \\ \hline

1 & $3.4^{\circ}$ & $6.6^{\circ}$ & $2.2 \times 10^{-3}$ & $< 2.6 \times 10^{-6}$ & $ > 7 \times 10^{5}$ \\

2 & $6.7^{\circ}$ & $8.8^{\circ}$ & $2.7 \times 10^{-4}$ & $< 1.5 \times 10^{-6}$ & $> 3 \times 10^{4}$ \\

3 & $11.2^{\circ}$ & $13.2^{\circ}$ & $6.6 \times 10^{-4}$ & $< 1.6 \times 10^{-6}$ & $> 1 \times 10^{5}$ \\ \hline

$1.6 \mu$m Imager \\ \hline

1 & $3.4^{\circ}$ & $6.6^{\circ}$ & $4.1 \times 10^{-3}$ & $< 3.0 \times 10^{-6}$ & $ > 1 \times 10^{6} $ \\

2 & $6.7^{\circ}$ & $8.8^{\circ}$ & $3.5 \times 10^{-4}$ & $< 1.9 \times 10^{-6}$ & $> 3 \times 10^{4}$ \\

3 & $11.2^{\circ}$ & $13.2^{\circ}$ & $1.3 \times 10^{-3}$ & $< 4.3 \times 10^{-6}$ & $> 9 \times 10^{4}$ \\

\hline

\end{tabular}
\label{tab:rings}
\end{table*}

Population 1 rings are generated by reflections off a lens mounting
flange (Figure \ref{fig:ringraytrace1}), and are produced by bright
sources between $3.4^{\circ}$ and $6.6^{\circ}$ off-axis. These rings
also have the strongest optical coupling, with an integrated flux in
the ring a few tenths of percent of the incident source flux. Given
their large acceptance angle, stars brighter than $4^{\mathrm{th}}$
magnitude are sufficiently abundant to generate multiple bright rings.

Following their discovery in the first flight data, we measured the
population 1 rings and searched for other optical reflections in the
laboratory.  We illuminated each Imager aperture with collimated light
and then scanned the angle of incidence of the collimated beam up to
$25^{\circ}$ off-axis.  The first set of measurements confirmed the
existence of the population 1 rings, and allowed the discovery of the
second class of fainter rings.

The second class of rings is comprised of two sub-populations which
are both generated by reflections off the lens tube and lens support
fixtures at the front of the optics assembly (Figure
\ref{fig:ringraytrace1}).  These rings have flux coupling coefficients
similar to, but slightly less than, the population 1 rings, but have
much larger solid angles on the array and so produce smaller per pixel
brightness.  Together, population 2 and 3 rings are caused by bright
sources $6.7^{\circ}$ to $13.2^{\circ}$ off-axis. These
rings are not readily visible in the images from the first flight,
though their presence was verified in the lab after flight.

Given the acceptance angles, star number counts and the quality of the
ancillary data, the first set of rings are sufficiently bright to be
modeled and masked from the first flight images.  However, the second set
of rings have a more complex morphology and fainter surface brightness,
and are more difficult for us to confidently account for in the images.

\begin{figure*}[htb]
\epsfig{file=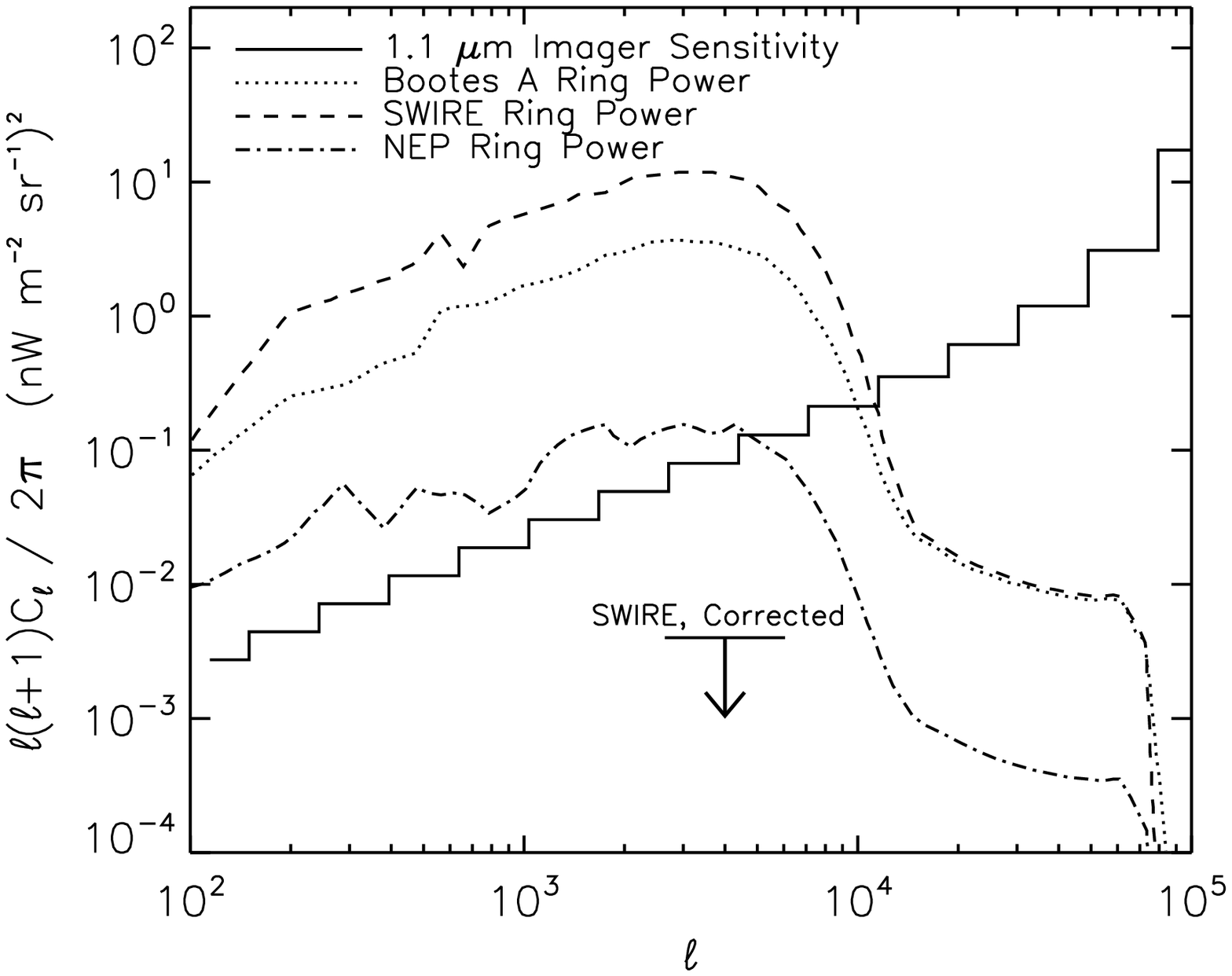,width=0.5\textwidth}
\epsfig{file=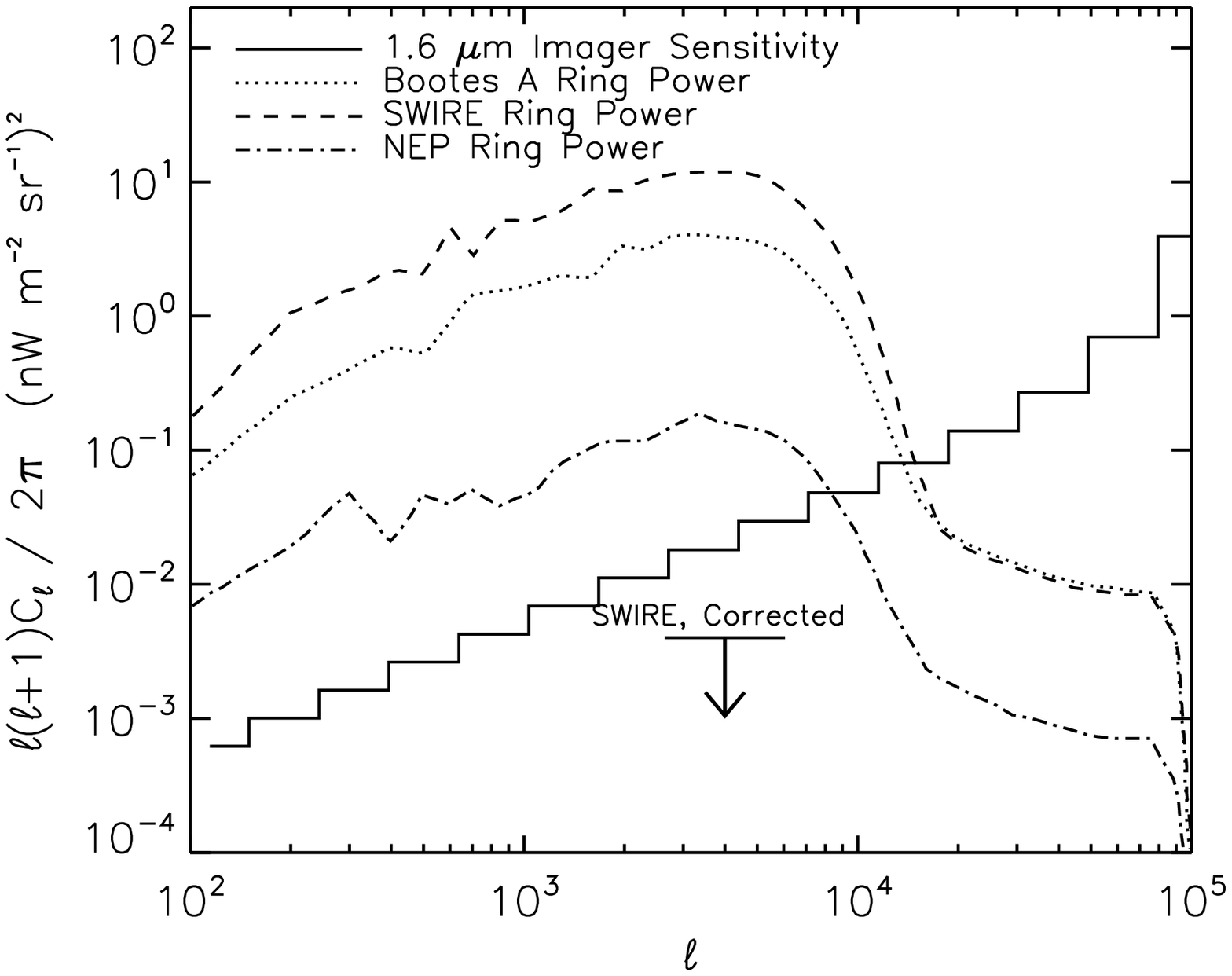,width=0.5\textwidth}
\caption{Simulated power spectra for the second class of rings for
  both Imager instruments, $1.1 \, \mu$m (left) and $1.6 \, \mu$m (right).
  These spectra were computed given the ring parameters in Table
  \ref{tab:rings} and the known star fluxes and positions near
  the CIBER fields.  The instrument sensitivity is the the same
  as modeled in Figure \ref{fig:pwrspec}.  The amplitude of the
  power spectrum of the rings is different for each field because
  of the differing stellar populations near each, but similar between
  the bands because of the typical color of stars.  For the second
  flight, the level of ring contamination is well below the instrument
  sensitivity, based on upper limits obtained in the laboratory following
  the modifications to the optics described in the text.  The upper limit
  is shown for SWIRE, the most demanding field.}
\label{fig:rings}
\end{figure*}

To understand the systematic error associated with the population 2
and 3 rings, we modeled their effect by convolving the measured
laboratory response with an off-axis star catalog for each field, and
calculated the spatial power spectrum of the resulting images.  These
rings, if left unmasked, produce power above the instrument
sensitivity level, as shown in Figure \ref{fig:rings}.

To remove the rings entirely, we made the optical simulation shown in
Figure \ref{fig:ringraytrace1}.  Following characterization of the
rings, the Imager optical assemblies were disassembled.  The
components responsible for the rings were grooved or cut back and
re-anodized.  The Imager optics were then reassembled, and the
off-axis measurements were repeated.  We did not observe any rings
following these modifications.  We place upper limits on the ring
coupling factors shown in Table \ref{tab:rings} which are based on the
uncertainty in the integrated surface brightness over the nominal ring
solid angles from the laboratory measurements.  We propagated these
upper limits through the model to produce synthetic images and then
power spectra.  The estimated reduction in the power spectrum from the
first class of rings are given in Table \ref{tab:rings}.  We find that
the effect on the power spectrum is negligible compared with the
instrument sensitivity after the optics modifications.

\section{Instrument Performance from the Second Flight}
\label{S:performance}

The Imagers were flown on the CIBER instrument on a second sounding
rocket flight in 2010 July.  All aspects of the experiment performed
well.  We found no evidence of bright thermal emission from the rocket
skin in either of the Imagers.  We did not observe rings in the flight images.
While the science data are still being analyzed, we summarize the observed
brightness and array photo-currents in Table \ref{tab:imagersens}.  Unfortunately,
it is difficult to estimate the full in-flight sensitivity in the power
spectrum without a noise estimator that accounts for correlated noise
in the presence of sources and masking.  Therefore we estimate the
in-flight per-pixel sensitivities by evaluating the noise in the
flight difference images (see Section \ref{sS:noise}).  The
corresponding per pixel surface brightness sensitivities, and point
source sensitivities using a $2 \times 2$ pixel aperture, are listed
in Table \ref{tab:imagersens}.  Our estimated sensitivity to the spatial
power spectrum is shown in Figure \ref{fig:newpssensitivity} based on the
variance of the power spectra of an ensemble of dark laboratory images
combined with flight photon noise.

\begin{table*}[ht]
\centering
\caption{Calculated and Second Flight Sensitivities in a $50 \,$s
  Observation.}
\begin{tabular}{lccccc}
\hline
 & \multicolumn{2}{c}{$1.1 \, \mu$m Imager} &
\multicolumn{2}{c}{$1.6 \, \mu$m Imager} \\

 & Predicted & Achieved & Predicted & Achieved \\ \hline

Sky brightness & 450 & 420 & 300 & 370 & nW m$^{-2}$ sr$^{-1}$\\

Photo current & 4.4 & 4.9 & 8.2 & 11.0 & \eps\\

Responsivity & 10 & 11 & 28 & 31 & me$^{-}$ s$^{-1}$ / nW m$^{-2}$ sr$^{-1}$\\

Current Noise & 0.31 & 0.35 & 0.41 & 0.45 & e$^{-}$ s$^{-1}$ ($ 1
  \sigma /$pix) \\

$\delta \lambda I_{\lambda}$ & 31.7 & 33.1 & 15.1 & 17.5  & nW
  m$^{-2}$ sr$^{-1}$ ($ 1 \sigma /$pix) \\

$\delta F_{\nu}$ & 18.5 & 18.4 & 18.2 & 17.8 & Vega Mag $(3 \sigma)$\\

\hline
\end{tabular}
\label{tab:imagersens}
\end{table*}

We scale the photo currents in Table \ref{tab:imagersens} to sky
brightness units using a calibration based on point sources observed
in flight.  We stacked sources with flux between 16.0 and
16.1 Vega magnitudes in the 2MASS catalog and integrated over the stacked
image to account for the extended PSF.  We converted this point source
calibration to surface brightness using the pixel solid angle, giving
the calibration factors in Table \ref{tab:imagersens}.

\section{Conclusions}

We have designed and tested an imaging instrument optimized to search
for the predicted spatial and spectral signatures of fluctuations from
the epoch of reionization.  The instrument demonstrates the sensitivity
needed to detect, or place interesting limits upon, REBL fluctuations
in the short observing time available in a sounding rocket flight.  We
have carried out a comprehensive laboratory characterization program
to confirm the focus, characterize the flat field response, perform an
end-to-end calibration, and measure the stray light response and
detailed noise properties.  After a first sounding
rocket flight in 2009 February, we modified the instrument to
eliminate response to thermal radiation from ambient portions of the
payload, and to reduce stray light to bright stars outside of the field
of view.  Scientific data from the second flight in 2010 July are currently
under analysis, and the instrument demonstrated sensitivity close to
design expectations.  The instrument characterization shows that systematic
errors from the extended PSF, stray light, and correlated noise over the
array are controlled sufficiently to allow a deep search for REBL spatial
fluctuations.  We recently completed a third flight in 2012 March that allows
us to cross-correlate images at different seasons to directly assess any ZL
fluctuations.  The flight and recovery were successful, and a fourth flight
is now planned.  A successor instrument, with 3 or more simultaneous spectral
bands and with higher sensitivity using a 30 cm telescope and improved
Hawaii-2RG arrays, is currently in development.

\section{Appendix}

The calculated sensitivities in Table \ref{tab:imagersens}, Figure \ref{fig:pwrspec}
and Figure \ref{fig:newpssensitivity} are based on a $50 \,$s integration with the
instrument parameters given in Table \ref{tab:imagerprops}.  The estimated photo
current $i_{\mathrm{phot}}$ given by:
\begin{equation}
\label{eq:Iph}
i_{\mathrm{phot}} \simeq \lambda I_{\lambda} \left( \frac{\eta A \Omega}{h \nu}
\frac{\Delta \lambda}{\lambda} \right) \hspace{0.5cm}
     [\mathrm{e}^{-}/\mathrm{s}],
\end{equation}
where $A \Omega$ is the pixel throughput, $\eta$ is the total
efficiency, $\lambda I_{\lambda}$ is the sky intensity, and $\Delta
\lambda$ is the integral bandwidth.  The term in brackets in Equation
\ref{eq:Iph} gives the surface brightness calibration from \eps\ to
\nw.  The current noise over an integration with continuous sampling
is given by:
\begin{equation}
\label{eq:deltaI}
\delta i_{\mathrm{phot}} = \sqrt{\frac{i_{\mathrm{phot}}}{T} + \delta
  Q_{\mathrm{CDS}}^{2}\frac{6 T_{0}}{T^{3}}} \hspace{0.5cm}
       [\mathrm{e}^{-}/\mathrm{s}],
\end{equation}
where $\delta Q_{\mathrm{CDS}}$ is the correlated double sample read
noise, $T = 50 \,$s is the integration time, and the frame rate
$T_{0}=1.78 \,$s.  The surface brightness sensitivity is therefore:
\begin{equation}
\label{eq:deltanuInu}
\delta \lambda I_{\lambda} = \delta i_{\mathrm{phot}} \frac{h \nu}{A \Omega \eta
  \Delta \lambda / \lambda} \hspace{0.5cm}  [\mathrm{nW \; m}^{-2} \;
  \mathrm{sr}^{-1}].
\end{equation}
Finally, the point source sensitivity is given by:
\begin{equation}
\label{eq:deltaF}
\delta \lambda F_{\lambda} = \delta i_{\mathrm{phot}} \frac{\sqrt{N_{pix}} h
  \nu}{A \eta \Delta \lambda / \lambda} \hspace{0.5cm}  [\mathrm{nW \;
    m}^{-2}],
\end{equation}
where $N_{pix}$ is the effective number of pixels that must be
combined to detect a point source, and we have assumed $N_{pix} = 4$.
These per-pixel sensitivities are used to estimate the sensitivity
on the power spectrum in Figure \ref{fig:pwrspec} and Figure \ref{fig:newpssensitivity}
using the formalism in \citet{Cooray2004}.  The calculation assumes
the noise in each pixel is independent, and ignores errors from source
removal and flat-field estimation.

\section*{Acknowledgments}

This work was supported by NASA APRA research grants NNX07AI54G,
NNG05WC18G, NNX07AG43G, NNX07AJ24G, and NNX10AE12G.  Initial support
was provided by an award to J.B.~from the Jet Propulsion Laboratory's
Director's Research and Development Fund.  Japanese participation in
CIBER was supported by KAKENHI (20$\cdot$34, 18204018, 19540250,
21340047 and 21111004) from Japan Society for the Promotion of Science
(JSPS) and the Ministry of Education, Culture, Sports, Science and
Technology (MEXT).  Korean participation in CIBER was supported by the
Pioneer Project from Korea Astronomy and Space science Institute
(KASI).

This publication makes use of data products from the Two Micron All
Sky Survey (2MASS), which is a joint project of the University of
Massachusetts and the Infrared Processing and Analysis
Center/California Institute of Technology, funded by the National
Aeronautics and Space Administration and the National Science
Foundation.  This work made use of images and/or data products
provided by the NOAO Deep Wide-Field Survey (NDWFS), which is
supported by the National Optical Astronomy Observatory, operated by
AURA, Inc., under a cooperative agreement with the National Science
Foundation.

We would like to acknowledge the dedicated efforts of the sounding
rocket staff at the NASA Wallops Flight Facility and the White Sands
Missile Range.  We also acknowledge the work of the Genesia
Corporation for technical support of the CIBER optics.  Our thanks to
Y.~Gong for sharing the REBL curves shown in Figure 1.
A.C.~acknowledges support from an NSF CAREER award, B.K.~acknowledges
support from a UCSD Hellman Faculty Fellowship, K.T.~acknowledges
support from the JSPS Research Fellowship for Young Scientists, and
M.Z.~acknowledges support from a NASA Postdoctoral Program Fellowship.

{\it Facility:} CIBER


\end{document}